\documentclass[article]{new-aiaa}
\usepackage[utf8]{inputenc}
\usepackage{textcomp}
\usepackage{xcolor}
\usepackage{graphicx}
\usepackage{amsmath}
\usepackage[version=4]{mhchem}
\usepackage{siunitx}
\usepackage{longtable,tabularx}
\usepackage{appendix}
\usepackage{multirow}
\usepackage{float}
\usepackage{tikz}
\usepackage{algorithm}
\usepackage{algpseudocode}

\usepackage{sectsty}
\usepackage{indentfirst}
\renewcommand\thesection{\arabic{section}}
\renewcommand\thesubsection{\thesection.\arabic{subsection}}
\usepackage{setspace}
\sectionfont{\fontsize{14}{12}\selectfont}
\subsectionfont{\fontsize{13}{12}\selectfont}
\subsubsectionfont{\fontsize{12}{12}\selectfont}
\newcommand*\circled[1]{\tikz[baseline=(char.base)]{
            \node[shape=circle,draw,inner sep=1pt] (char) {#1};}}

\providecommand{\keywords}[1]
{
  \small	
  \textbf{\textit{Keywords---}} #1
}

\title{Assessing Coronary Microvascular Dysfunction using Angiography-based Data-driven Methods}
\author{\fontsize{10pt}{14pt}\selectfont Haizhou Yang\textsuperscript{1}, Jiyang Zhang\textsuperscript{1,2}, Brahmajee K. Nallamothu\textsuperscript{3}, Krishna Garikipati\textsuperscript{4} and C. Alberto Figueroa\textsuperscript{1,5}\footnote{Corresponding author. E-mail address: figueroc@med.umich.edu}}

\affil{\fontsize{10pt}{14pt}\selectfont \textsuperscript{1}Department of Biomedical Engineering, University of Michigan, Ann Arbor, MI 48109, USA}
\affil{\fontsize{10pt}{14pt}\selectfont \textsuperscript{2}Department of Mechanical Science and Engineering, Sichuan University, Chengdu, Sichuan 610017, China}
\affil{\fontsize{10pt}{14pt}\selectfont \textsuperscript{3}Department of Internal Medicine, University of Michigan, Ann Arbor, MI 48109, USA}
\affil{\fontsize{10pt}{14pt}\selectfont \textsuperscript{4}Department of Aerospace and Mechanical Engineering, University of Southern California, Los Angeles, CA 90089, USA}
\affil{\fontsize{10pt}{14pt}\selectfont \textsuperscript{5}Department of Surgery, University of Michigan, Ann Arbor, MI 48109, USA}
\begin{document}

\maketitle

\begin{abstract}

Coronary microvascular dysfunction (CMD), characterized by impaired regulation of blood flow in the coronary microcirculation, plays a key role in the pathogenesis of ischemic heart disease and is increasingly recognized as a contributor to adverse cardiovascular outcomes. Despite its clinical importance, CMD remains underdiagnosed due to the reliance on invasive procedures such as pressure wire-based measurements of the index of microcirculatory resistance (IMR) and coronary flow reserve (CFR), which are costly, time-consuming, and carry procedural risks. To date, no study has sought to quantify CMD indices using data-driven approaches while leveraging the rich information contained in coronary angiograms. To address these limitations, this study proposes a novel data-driven framework for inference of CMD indices based on coronary angiography. A physiologically validated multi-physics model was used to generate synthetic datasets for data-driven model training, consisting of CMD indices and computational angiograms with corresponding contrast intensity profiles (CIPs). Two neural network architectures were developed: a single-input-channel encoder–MLP model for IMR prediction and a dual-input-channel encoder–MLP model for CFR prediction, both incorporating epistemic uncertainty estimation to quantify prediction confidence. Results demonstrate that the data-driven models achieve high predictive accuracy when evaluated against physics-based synthetic datasets, and that the uncertainty estimates are positively correlated with prediction errors. Furthermore, the utility of CIPs as informative surrogates for coronary physiology is demonstrated, underscoring the potential of the proposed framework to enable accurate, real-time, image-based CMD assessment using routine angiography without the need for more invasive approaches.

\end{abstract}


\section{Introduction}

More than 90\% of the total flow resistance in the coronary circulation originates from pre-arterioles, arterioles, and capillaries within the microvasculature \cite{geng2022index,vancheri2020coronary}. Consequently, microcirculation plays an important role in the regulation of coronary blood flow, ensuring a balance between oxygen and nutrient supply and metabolic demand. Coronary microvascular dysfunction (CMD), a significant concern in cardiology, is characterized by impaired blood flow and dysregulation within the microcirculation \cite{vancheri2020coronary,komaru2000coronary}. Structural abnormalities associated with CMD may arise from atherosclerosis, vascular remodeling, or fibrosis, whereas functional abnormalities can result from coronary vasospasm, endothelial cell and smooth muscle dysfunction, or metabolic derangements \cite{taqueti2018coronary}. These abnormalities disrupt the delicate balance between myocardial perfusion and demand, leading to severe clinical outcomes, including heart failure, myocardial infarction, stroke, and increased mortality \cite{camici2007coronary,godo2021coronary}.

Diagnostic modalities to evaluate the functional state of CMD \cite{sucato2022classification} include thrombolysis in myocardial infarction (TIMI) flow grade and TIMI frame count derived from coronary angiography, perfusion assessment using positron emission tomography (PET), and, less frequently, catheter-based techniques \cite{lanza2010primary}. Coronary angiography, an X-ray-based imaging technique, is widely used in clinical practice, with millions of procedures performed annually worldwide. It provides high-resolution imaging of contrast agent dynamics within the coronary arteries, offering valuable insights into vascular morphology, motion, and deformation. TIMI flow grade and TIMI frame count are two angiography-based methods developed for CMD assessment, utilizing contrast washout patterns assessed visually by operators to evaluate microvascular perfusion and coronary flow dynamics \cite{doherty2021predictors}. Although these methods enable a rapid assessment of microvascular perfusion based on flow characteristics, they only provide qualitative and somewhat subjective and limited information, making them rarely used in practice. Myocardial positron emission tomography (PET) is an advanced imaging modality that provides high-resolution images and accurate measurements of myocardial blood flow at both rest and pharmacological stress, facilitating the detection of ischemia and microvascular dysfunction \cite{blankstein2014cardiac,el2020myocardial}. However, despite its diagnostic accuracy and prognostic value, the widespread adoption of PET remains limited due to high operational costs, the need for specialized equipment, and restricted availability to select medical centers. 

Recently, invasive wire-based techniques, such as the index of microcirculatory resistance (IMR) and coronary flow reserve (CFR), have gained increasing attention as widely adopted invasive methods for assessing coronary microcirculatory function \cite{fearon2003novel,ford2017coronary,martinez2015index,gulati20212021}. The assessment of these CMD indices involves the insertion of a coronary wire to simultaneously measure pressure and estimate flow. IMR is a quantitative measure of microvascular resistance, calculated using distal coronary pressure and hyperemic mean transit time to estimate the flow, providing an index of microvascular dysfunction independent of epicardial disease. CFR, on the other hand, represents the ratio of maximal hyperemic to resting coronary blood flow, reflecting both epicardial and microvascular contributions to coronary circulation. Notably, wire-based assessments have been shown to reliably predict myocardial viability following primary angioplasty for myocardial infarction and to determine the extent and severity of myocardial infarction in affected patients \cite{fearon2004microvascular,mcgeoch2010index,lim2009usefulness,fearon2008predictive,clarke2020invasive}. However, despite their clinical significance, these techniques remain underutilized due to their invasive nature, procedural complexity, and associated patient risks. Consequently, there is a pressing demand for a non-invasive computational approach to replicate these wire-based measurements.  

Several studies have developed computational fluid dynamics (CFD)-based methods to compute the IMR \cite{jiang2022novel,huang2023coronary,ai2020coronary} and CFR \cite{morris2021novel} using geometries reconstructed from angiographic data. However, these approaches face significant challenges related to the setup of boundary conditions in CFD models. In many cases, either invasive fractional flow reserve (FFR) is employed to provide pressure boundary conditions, or TIMI frame count is used to estimate flow boundary conditions. Invasive FFR, similar to invasive IMR or CFR assessment, is costly and poses additional risks to patients. On the other hand, flow estimation via TIMI frame count lacks accuracy. Moreover, the high computational cost associated with CFD simulations leads to prolonged processing times and increased demand for computational resources. Yong et al. introduced a novel method for IMR calculation in the presence of epicardial stenosis by establishing a linear regression model that relates coronary FFR to myocardial FFR \cite{yong2013calculation}. However, this approach could be enhanced by employing a more sophisticated modeling technique, and it still relies on high-risk, invasive FFR measurements. In an effort to reduce invasiveness, Tebaldi et al. proposed replacing FFR measurements with the quantitative flow ratio (QFR) \cite{tebaldi2020angio}. Nevertheless, QFR estimation requires CFD simulations, and inaccuracies in boundary condition setup may compromise its reliability.  

In summary, we submit that there are two major challenges for non-invasively assessing CMD. First, existing methods fail to fully leverage the rich dynamic information embedded in coronary angiography. Current angiography-based approaches primarily rely on static or semi-quantitative metrics, overlooking valuable flow dynamics that could enhance diagnostic accuracy. Second, available methods either involve highly invasive procedures, such as wire-based measurements, or suffer from limitations in computational efficiency and accuracy, as seen in CFD-based approaches. Given these challenges, there is a pressing need to develop accurate and computationally efficient data-driven methods to assess CMD using dynamic coronary angiography. 

In this study, a data-driven framework capable of harnessing dynamics encapsulated within coronary angiography is developed to quantify both IMR and CFR. In contrast to existing research efforts, the proposed framework introduces several key innovations: (1) To the best of our knowledge, this is the first effort to leverage the dynamic information in coronary angiography for predicting CMD indices, including IMR and CFR. Data-driven models are employed to establish the relationship between coronary angiography dynamics and CMD indices. (2) The framework incorporates uncertainty quantification in the predictions, providing confidence intervals for estimated CMD indices. These confidence intervals offer critical insights for clinical decision-making, enhancing diagnostic reliability, and assisting in risk stratification.  

The rest of this paper is organized as follows. Section \ref{Methodology} introduces the proposed data-driven framework, focusing on a multi-physics CFD model used to create synthetic angiograms, simulation data generation and results extraction, design of data-driven models, and uncertainty quantification. Sections \ref{Results} and \ref{Discussion} present the results and corresponding discussions. Finally, Section \ref{Conclusion} provides the conclusions and outlines directions for future research.

\section{Methodology}\label{Methodology}
\subsection{Framework Overview}

\begin{figure}[h]
\centering
\includegraphics[width=1\textwidth]{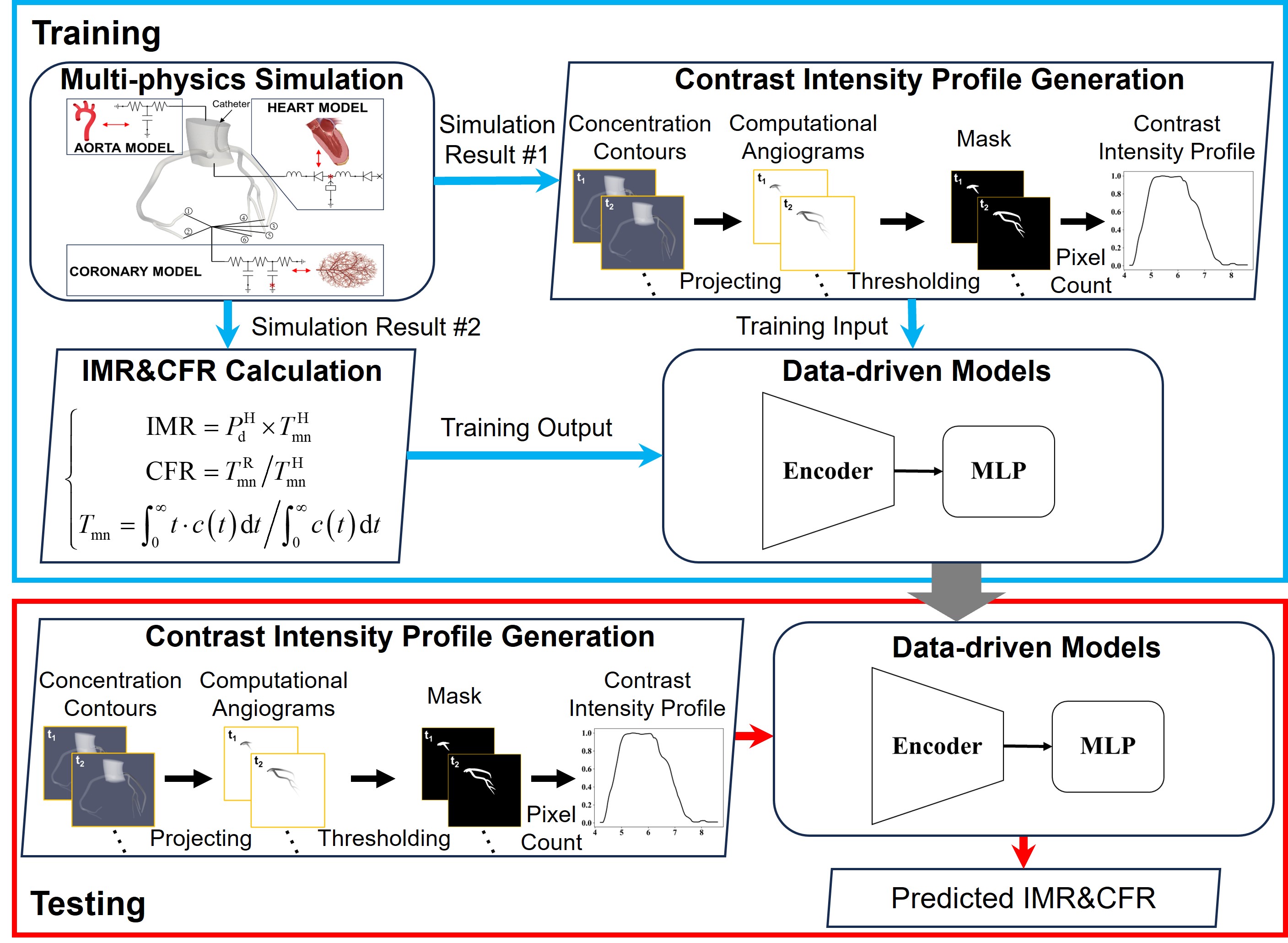}
\caption{\centering ML-IMR framework, including offline model generation and online prediction components.}
\label{fig:Framework}
\end{figure}

The proposed data-driven framework for IMR and CFR assessment consists of two stages, as shown in Fig. \ref{fig:Framework}: training and testing (inference). 

The training stage (blue box in Fig. \ref{fig:Framework}) consists of four key components: (1) multi-physics CFD simulation; (2) extraction of simulation results; and (3) development of data-driven models. First, the multi-physics CFD model of contrast injection, previously developed in \cite{yang2024multi}, is used to generate computational angiograms, by simultaneously solving the Navier-Stokes equations and transport of a scalar quantity (i.e., the contrast agent) through the coronary arteries.  These simulations are conducted under both hyperemic and resting conditions, as CMD indices are defined based on both states. This model employs a well-established 3D-0D modeling approach, where the 3D component captures blood flow dynamics in the epicardial coronary arteries, whereas lumped parameter models (LPMs) are used to represent coronary microcirculation. This multi-physics model generates physiologically relevant hemodynamic parameters, including pressure and flow patterns, as well as contrast injection and washout. Next, two types of simulation results are extracted (detailed in Section \ref{Result_Extraction}): (1) CIPs from contrast concentration contours through projection, thresholding, and pixel counting, as seen in previous studies \cite{yang2024multi,resnick2024neural} and (2) CMD indices (IMR and CFR) derived from pressure and flow results. Finally, the extracted CIPs and corresponding CMD indices are used as paired input-output data pairs to train data-driven models, establishing relationships between contrast dynamics and CMD indices.

During the testing (inference) stage (red box in Fig. \ref{fig:Framework}), the trained data-driven models are applied to previously unseen CIPs to estimate CMD indices. For IMR prediction, which is defined exclusively under hyperemic conditions, only hyperemic CIPs are used as model input. In contrast, CFR prediction, by definition, requires CIPs from both resting and hyperemic states to generate accurate estimates.

\subsection{Multi-physics CFD Model}

\begin{figure}[h]
\centering
\includegraphics[width=0.8\textwidth]{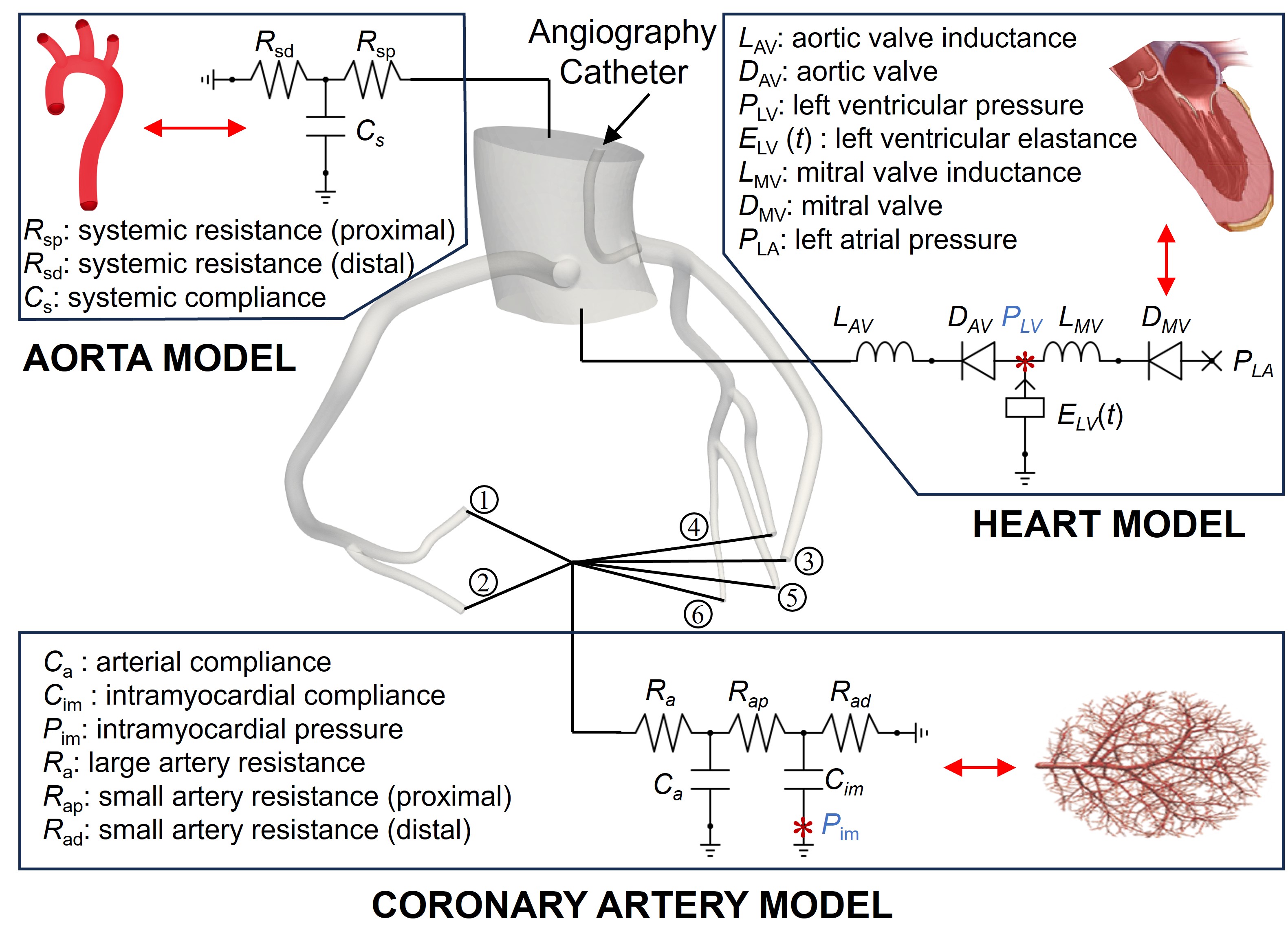}
\caption{\centering 3D-0D Multi-physics model of contrast injection.}
\label{fig:CFD_model}
\end{figure}

A multi-physics CFD model for iodine contrast injection was developed using CRIMSON \cite{arthurs2021crimson}. This model simultaneously solves for both hemodynamics (e.g., blood flow) and the advection-diffusion transport of a scalar quantity (e.g., the contrast agent). The complete formulation of the multi-physics CFD model was previously reported in \cite{yang2024multi}; here, we provide a brief summary for completeness. A schematic representation of the multi-physics model is shown in Fig. \ref{fig:CFD_model}. The anatomical model consists of the aortic root and the primary branches of both the left and right coronary arteries, reconstructed from coronary computed tomography angiography (CCTA) data of a 64-year-old female patient with suspected CMD. The right coronary tree includes the right coronary artery (\circled{1} RCA) and the acute marginal (\circled{2} AM) branches. The left coronary arterial tree comprises four branches: the left anterior descending (\circled{3} LAD), obtuse marginal (\circled{4} OM1, \circled{5} OM2), and left circumflex (\circled{6} LCx) arteries. Additionally, the model incorporates the distal segment of an angiography catheter to simulate contrast agent injection.

\underline{Governing equations:} The governing equations for blood flow and contrast transport are solved using the finite element method within the computational domain ($\Omega$). Specifically, the multi-physics model solves the incompressible Navier–Stokes equations for velocity $\mathbf{u}$ and pressure $p$, along with the advection-diffusion equation for the contrast agent concentration $c$:

\begin{equation}
\label{eqn:NS}
\begin{aligned}
&   \left\{
\begin{aligned}
&\nabla \cdot \mathbf{u} = 0 \\
&\frac{\partial \mathbf{u}}{\partial t} + (\mathbf{u} \cdot \nabla) \mathbf{u} = -\frac{1}{\rho} \nabla p + \nu \nabla^2 \mathbf{u}
\end{aligned}
\right. \quad\text{on}\quad \Omega,\\
\end{aligned}
\end{equation}

\begin{equation}
\label{eqn:CD}
\frac{\partial c}{\partial t} + \nabla \cdot (\mathbf{u} c) = D \nabla^2 c\quad\text{on}\quad \Omega,
\end{equation}
where $\rho$ denotes blood density, $\nu$ blood kinematic viscosity, and $D$ the diffusion coefficient of the scalar $c$ in blood.

\underline{Lumped parameter models (LPMs):} To capture physiologically relevant pulsatile coronary and aortic hemodynamics, the 3D computational domain is coupled to lumped parameter models (LPMs) \cite{arthurs2016mathematical,kim2010patient,westerhof2009arterial}. Three LPMs are incorporated: (1) an aortic model at the aortic outlet, (2) a heart model describing ventricular contraction at the aortic inlet, and (3) coronary models at the outlets of the six coronary branches.

\begin{figure}[H]
\centering
\includegraphics[width=0.4\textwidth]{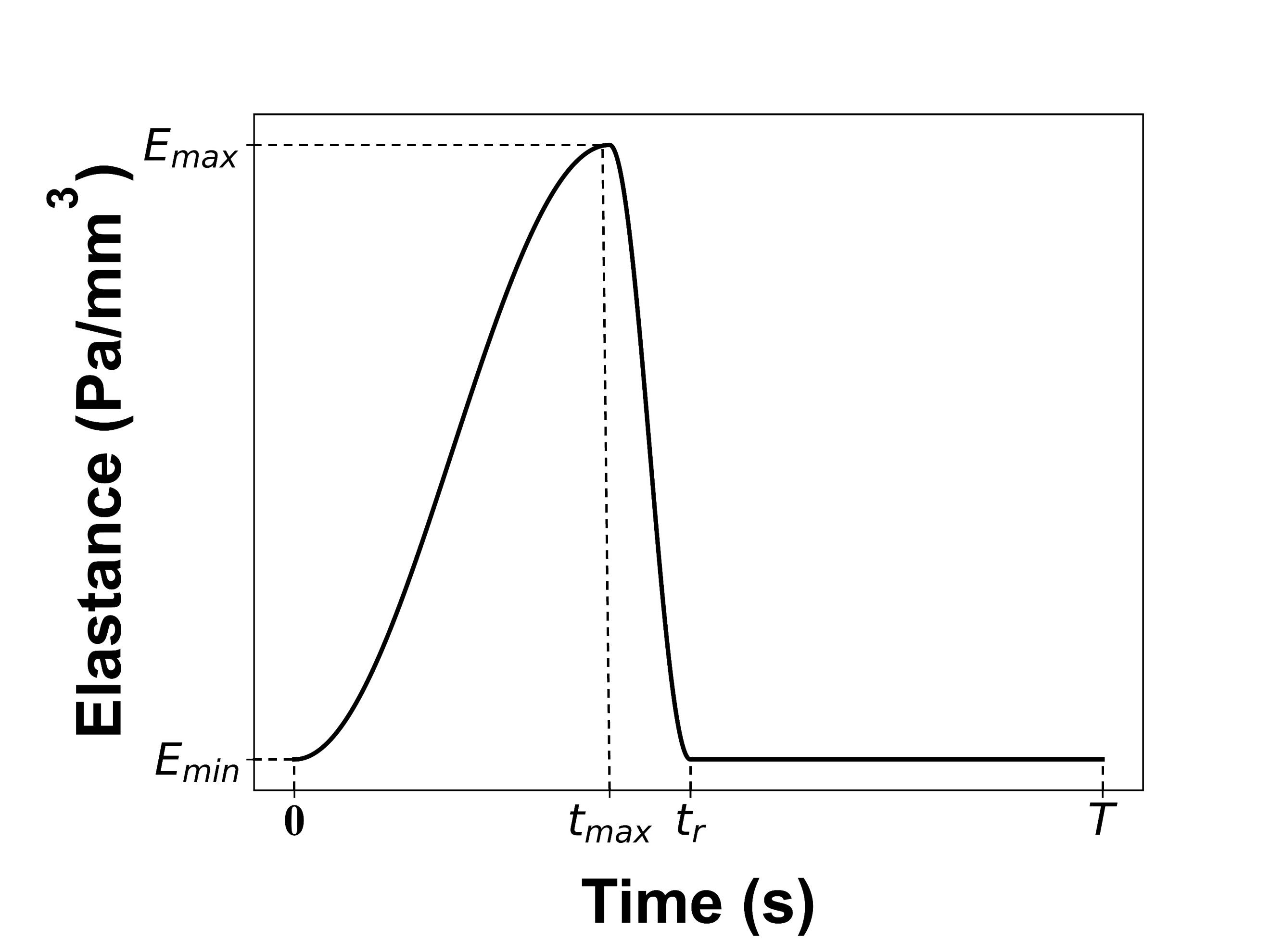}
\caption{\centering Elastance function.}
\label{fig:elastance}
\end{figure}

The heart model consists of an aortic valve inductance $L_\text{AV}$, an aortic valve $D_\text{AV}$, a time-varying left ventricular elastance function $E_\text{LV}(t)$, a mitral valve inductance $L_\text{MV}$, a mitral valve $D_\text{MV}$, and a fixed left atrial pressure $P_\text{LA}$. Both aortic and mitral valves are represented by diodes and assume perfect function (i.e., lack of regurgitant flow). The elastance function $E_\text{LV}(t)$ represents ventricular contractility \cite{pope2008estimation} and is parameterized by five variables, as shown in Fig. \ref{fig:elastance}: maximum elastance $E_\text{max}$, minimum elastance $E_\text{min}$, cardiac cycle duration $T$, time to maximum elastance $t_\text{max}$, and relaxation time $t_\text{r}$. 

The coronary LPMs account for myocardial compression effects, producing characteristic diastolic-dominated coronary flow waveforms \cite{kim2010patient}. Each coronary LPM consists of a large artery resistance $R_\text{a}$, a proximal small artery resistance $R_\text{ap}$, a distal small artery resistance $R_\text{ad}$, an arterial compliance $C_\text{a}$, and an intramyocardial compliance $C_\text{im}$.

The aortic outlet is connected to an aorta LPM, defined via a three-element Windkessel model \cite{westerhof2009arterial}, consisting of systemic compliance $C_\text{s}$, proximal systemic resistance $R_\text{sp}$, and distal systemic resistance $R_\text{sd}$.

\begin{figure}[h]
\centering
\includegraphics[width=0.4\textwidth]{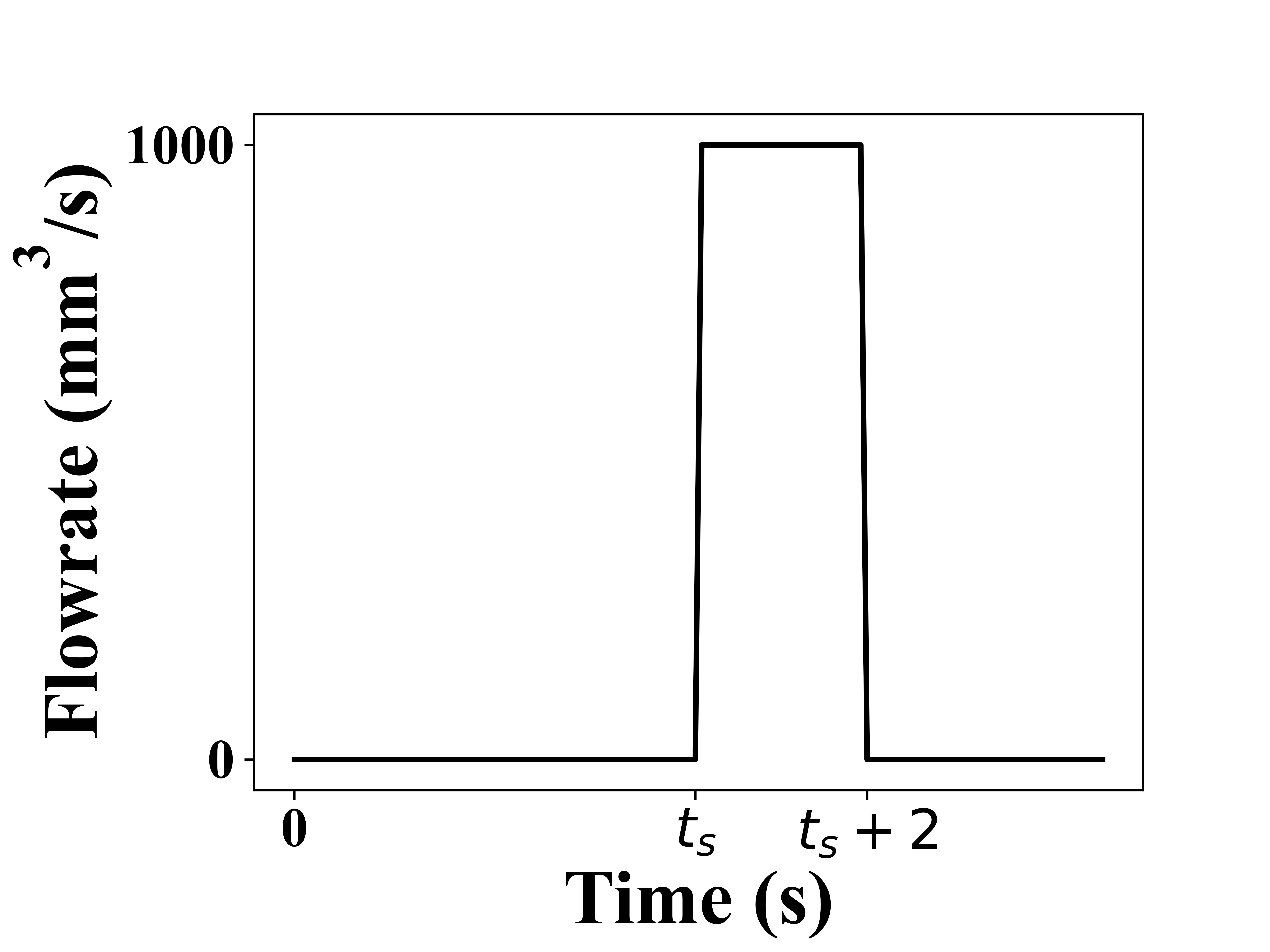}
\caption{\centering Profiles of contrast agent injection at catheter inlet.}
\label{fig:Catheter_flow}
\end{figure}

\underline{Contrast injection parameters:} Contrast injection is modeled as a step function, mimicking clinical procedures, as shown in Fig. \ref{fig:Catheter_flow}. The injected contrast volume is set to 2 milliliters, consistent with clinical practice \cite{sacha2019ultra}, with a concentration of $c_{0} = 400$ mg/mL and a diffusivity of $D = 0.00203$ $\text{mm}^2/\text{s}$ \cite{caschera2016contrast}. Injection begins at $t = t_\text{s}$ and proceeds at a constant rate of 1000 $\text{mm}^3/\text{s}$ over 2 s.

\underline{Simulation setup:} Blood was modeled as a Newtonian fluid with viscosity $\mu = 0.004$ Pa·s and density $\rho = 1060$ $\mathrm{kg/m}^3$. The heart rates for resting and hyperemic conditions were set to 60 bpm (cycle duration = 1 s) and 82 bpm (cycle duration = 0.73 s), respectively. Resting-state simulations were conducted over 8 cardiac cycles (total duration = 8 s), with the first two cycles used to establish periodic hemodynamics before contrast injection in the third cycle. Hyperemic-state simulations were run for 9 cardiac cycles (6.57 s), with contrast injected during the fifth cycle. A time step size of 0.0001 s and a convergence tolerance of $1\text{e}{-3}$ were used. The computational domain was discretized using an anisotropic and curvature-based mesh refinement strategy, yielding approximately 1.5 million linear tetrahedral elements after mesh-independence analysis. Simulations were performed using the CRIMSON solver \cite{arthurs2021crimson} on 108 computing cores, requiring approximately 50 hours per case.

\subsection{Simulation Data Generation}
\label{DataGeneration}
Multi-physics simulations were conducted under both resting and hyperemic conditions to generate hemodynamic data, computational CIPs and computational CMD indices, for data-driven model training. This process involves several key steps.

First, parameters for two reference healthy hemodynamic conditions (one at rest, the other at hyperemia) were calibrated using methods outlined in our previous study \cite{yang2024multi}. Briefly, a two-stage calibration process was employed: the first stage tuned heart and aortic LPM parameters (systemic compliance $C_\text{s}$, proximal systemic resistance $R_\text{sp}$, distal systemic resistance $R_\text{sd}$, maximum elastance $E_\text{max}$, minimum elastance $E_\text{min}$, time to maximum elastance $t_\text{max}$, and left atrial pressure $P_\text{LA}$) to match systemic hemodynamic targets, while the second stage adjusted coronary LPM parameters (large artery resistance $R_{\text{a},i}$, proximal small artery resistance $R_{\text{ap},i}$, distal small artery resistance $R_{\text{ad},i}$, arterial compliance $C_{\text{a},i}$, and intramyocardial compliance $C_{\text{im},i}$, for $i = 1,2,...,6$) to reproduce CIPs and invasive CFR measurements. Calibrated LPM parameters are detailed in Tables \ref{tab:Heart_parameters}, \ref{tab:Optimized_Heart_parameters}, and \ref{tab:coronary_parameters}. Specifically, Tables \ref{tab:Heart_parameters} and \ref{tab:Optimized_Heart_parameters} provide parameters for heart and aortic LPMs, while Table \ref{tab:coronary_parameters} lists coronary LPM parameters.

\begin{table}[H]
\centering
\footnotesize
\renewcommand{\arraystretch}{1.5} 
\setlength{\tabcolsep}{5pt}
\caption{\centering Parameters of heart LPM and elastance function, which are not part of the calibration process.}
\label{tab:Heart_parameters}
\begin{tabular}{c c c c c c c}

\hline
  $R_{\text{MV}}$ & $L_{\text{MV}}$ & $R_{\text{AV}}$ & $L_{\text{AV}}$ & $t_\text{r} - t_\text{max}$ & $T_{\text{Rest}}$ & $T_{\text{Hyper}}$ \\ 
  ($\text{Pa} \cdot \text{s} \cdot \text{mm}^{-3}$) & ($\text{Pa} \cdot \text{s}^2 \cdot \text{mm}^{-3}$) & ($\text{Pa} \cdot \text{s} \cdot \text{mm}^{-3}$) & ($\text{Pa} \cdot \text{s}^2 \cdot \text{mm}^{-3}$) & (s) & (s) & (s) \\
\hline
  $3.9 \times 10^{-4}$ & $1 \times 10^{-5}$ & $1 \times 10^{-5}$ & $1 \times 10^{-5}$ & 0.1 & 1 & 0.73 \\
\hline
\end{tabular}
\end{table}

\begin{table}[H]
\centering
\footnotesize

\renewcommand{\arraystretch}{1.5} 
\setlength{\tabcolsep}{5pt}
\caption{\centering Calibrated parameters of aortic and heart LPMs for the reference resting and hyperemic conditions.}
\label{tab:Optimized_Heart_parameters}
\resizebox{\textwidth}{!}{  
\begin{tabular}{c c c c c c c c}
\hline
  & $C_\text{s}$ & $R_\text{sp}$ & $R_\text{sd}$ & $E_\text{max}$ & $E_\text{min}$ & $t_\text{max}$ & $P_\text{LA}$  \\ 
  & ($\text{mm}^3 \cdot \text{Pa}^{-1}$) & ($\text{Pa} \cdot \text{s} \cdot \text{mm}^{-3}$) & ($\text{Pa} \cdot \text{s} \cdot \text{mm}^{-3}$) & ($\text{Pa} \cdot \text{mm}^{-3}$) & ($\text{Pa} \cdot \text{mm}^{-3}$) & (s) & (Pa)  \\
\hline

Rest& 18.382& 0.009& 0.158& 0.190& 0.015& 0.390& 2286.880
\\ 

Hyperemia& 16.361& 0.005& 0.087& 0.228& 0.015& 0.409& 2376.910
\\ 

\hline
\end{tabular}
}
\end{table}

\begin{table}[H]
\centering
\footnotesize
\renewcommand{\arraystretch}{1.5} 
\setlength{\tabcolsep}{5pt}
\caption{\centering Calibrated parameter values in coronary LPMs for the reference resting and hyperemic conditions.}
\label{tab:coronary_parameters}
\begin{tabular}{c c c c c c c}
\hline
  &  & $R_\text{a}$ & $R_\text{ap}$ & $R_\text{ad}$ & $C_\text{a}$ & $C_\text{im}$ \\ 
  &  & ($\text{Pa} \cdot \text{s} \cdot \text{mm}^{-3}$) & ($\text{Pa} \cdot \text{s} \cdot \text{mm}^{-3}$) & ($\text{Pa} \cdot \text{s} \cdot \text{mm}^{-3}$) & ($\text{mm}^3 \cdot \text{Pa}^{-1}$) & ($\text{mm}^3 \cdot \text{Pa}^{-1}$) \\
\hline
     \multirow{6}{*}{Rest}& LAD &4.544& 1.363& 12.696& 0.014& 0.135
\\ 
     & OM1 &3.732& 1.120& 10.429& 0.014& 0.135
\\ 
     & OM2 &7.153& 2.146& 19.989& 0.007& 0.135
\\
     & LCx &6.398& 1.919& 17.878& 0.007& 0.135
\\
     & AM &4.757& 1.427& 13.293& 0.012& 0.189
\\
     & RCA &3.199& 0.960& 8.939& 0.012& 0.189
\\ 
\hline
     \multirow{6}{*}{Hyperemia}& LAD &1.159& 0.348& 3.239& 0.014& 0.128
\\ 
     & OM1 &0.952& 0.286& 2.660& 0.014& 0.128
\\
     & OM2 &1.825& 0.547& 5.099& 0.007& 0.128
\\
     & LCx &1.632& 0.490& 4.561& 0.007& 0.128
\\
     & AM &1.214& 0.364& 3.391& 0.011& 0.180
\\
     & RCA &0.816& 0.245& 2.280& 0.011& 0.180
\\
     
\hline
\end{tabular}
\end{table}

Second, starting from each reference hemodynamic condition, multiple simulations are performed by modifying the coronary LPMs to represent patients with different CMD severities. Additionally, the contrast injection profile is adjusted to simulate different injection timings. To characterize the numerous permutations of coronary LPM parameters and timing of contrast injection, we defined a total of four design variables $x_{i}, i=1,...,4$: $x_1$ simultaneously scales the magnitude of $R_\text{a}$ and $R_\text{ap}$; $x_2$ scales $R_\text{ad}$; $x_3$, simultaneously scales $C_\text{a}$ and $C_\text{im}$; $x_4$ modifies the injection timing within the cardiac cycle.

\begin{figure}[h]
\centering
\includegraphics[width=0.7\textwidth]{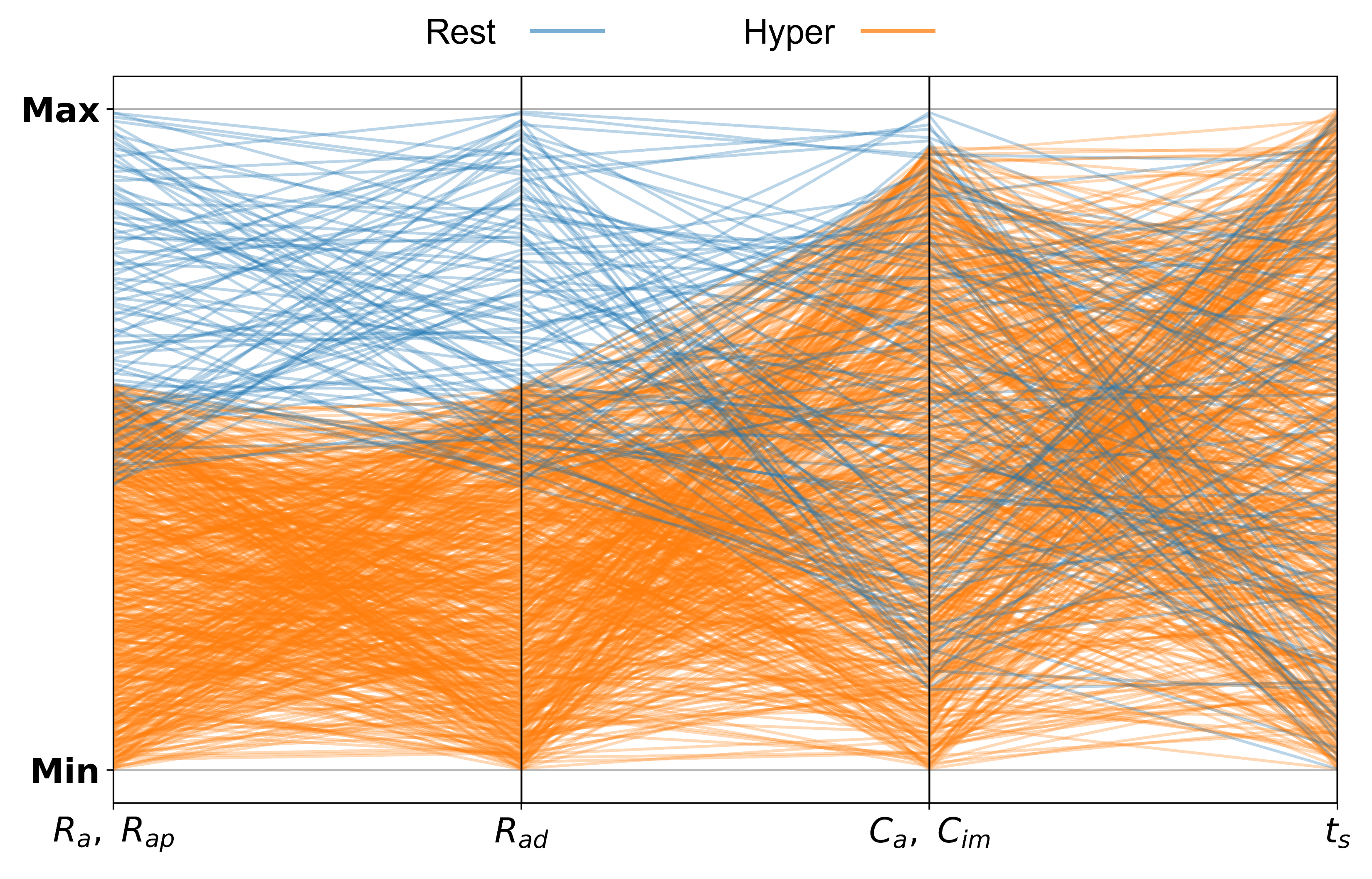}
\caption{\centering Parallel coordinates plot illustrating the distribution of sampled design variables for both hyperemic and resting conditions.}
\label{fig:parallel_coordinates}
\end{figure}

The ranges of these design variables are listed in Table \ref{tab:range_DV}. To simplify our analysis, we assumed that all coronary arteries exhibit a similar degree of CMD, implying that each corresponding LPM parameter for the different vessels is scaled by the same design variable. Note that the range under resting conditions is narrower than under hyperemia, reflecting the reduced hemodynamic variability among patients at rest compared to hyperemia. A total of 600 simulations were conducted under hyperemic conditions and 100 under resting conditions, using Latin Hypercube Sampling (LHS) within their respective design variable ranges. Fig.~\ref{fig:parallel_coordinates} illustrates the distribution of simulation parameters, where each reference value is scaled by the corresponding design variable, i.e. $R_\text{a} = R_\text{a}^{ref} \cdot x_1$, using parallel coordinates plots for both hyperemic and resting conditions, highlighting the coverage of the sampled parameter space. Because the reference resistance and capacitance values are different between resting and hyperemic conditions, their corresponding ranges exhibit only limited overlap.

\begin{table}[H]
\centering
\footnotesize
\renewcommand{\arraystretch}{1.5} 
\setlength{\tabcolsep}{5pt}
\caption{\centering Range of design variables for the simulations under resting and hyperemic conditions.}
\label{tab:range_DV}
\begin{tabular}{c c c c c}
\hline
  & $x_1(R_\text{a} \& R_\text{ap})$ & $x_2(R_\text{ad})$ & $x_3(C_\text{a}\&C_\text{im})$ & $x_4(t_\text{s})$  \\
\hline

Rest& [1,2]& [1,2]& [0.5,1]& [0,1]\\ 

Hyperemia& [1,5]& [1,5]& [0.1,1]& [0,1]\\ 

\hline
\end{tabular}
\end{table}

\subsection{Simulation Results Extraction} \label{Result_Extraction}

Once the 600 hyperemic and 100 resting simulations were completed, relevant information was extracted from the results to train data-driven models. This included CIPs and a \textit{mean transit time} for vessel $i$ ($T_{\text{mn},i}$), which is used to compute CMD indices ($\text{CFR}_{i}$ and $\text{IMR}_{i}$). 

The workflow for CIP extraction was previously described in \cite{yang2024multi}. For completeness, we provide a brief summary here. From the multi-physics CFD simulations, the 3D iodine contrast concentration field $c$ is projected onto a 2D plane using a view angle (RAO = $21.9^\circ$, CAU = $18.3^\circ$) consistent with clinical X-ray angiograms to create time-series of computational angiograms (see \textit{Contrast Intensity Profile Generation} block of Fig. \ref{fig:Framework}). These images are then binarized via thresholding to generate segmentation masks. The number of bright pixels in each frame, corresponding to the area occupied by iodine contrast, is counted, and a normalized CIP is obtained by plotting the normalized pixel count over time. 

In the clinical setting, when measuring CMD indices, coronary blood flow is indirectly assessed using a thermodilution technique \cite{candreva2021basics}. This method involves injecting a bolus of room-temperature saline into the coronary artery, and measuring the resulting temperature changes downstream using a pressure-temperature sensor-tipped guidewire. The mean transit time of the saline bolus is inversely related to blood flow. In our multi-physics simulations, the mean transit time was estimated by tracking the downstream changes of contrast agent concentration $c$, leveraging the fact that both temperature and concentration are governed by an advection-diffusion equation. The mean transit time for vessel $i$ ($T_{\text{mn},i}$) was computed as:
\begin{equation}
\label{eqn:transit_time}
T_{\text{mn},i} = \frac{\int_0^\infty t \cdot c_i(t) \, dt}{\int_0^\infty c_i(t) \, dt},
\end{equation}
where $c_i(t)$ denotes the time-dependent contrast agent concentration at the measurement location for vessel $i$. Since $t$ in the numerator's integrand is inversely proportional to the flow, the transit time $T_{\text{mn},i}$ is also inversely proportional to flow. Finally, CMD indices for vessel $i$ can be computed as:

\begin{equation}
\label{eqn:imr_i}
\text{IMR}_{i} = p_{\text{d},i}^\text{H} \times T_{\text{mn},i}^\text{H},
\end{equation}

\begin{equation}
\label{eqn:cfr_i}
\text{CFR}_{i} = \frac{T_{\text{mn},i}^\text{R}}{T_{\text{mn},i}^\text{H}},
\end{equation}
where $p_{\text{d},i}^\text{H}$ is the distal coronary pressure measured under hyperemic conditions for vessel $i$, and $T_{\text{mn},i}^\text{R}$ and $T_{\text{mn},i}^\text{H}$ are the mean transit times for vessel $i$ at rest and during hyperemia, respectively.

Equations~(\ref{eqn:transit_time})-(\ref{eqn:cfr_i}) reflect the manner in which CMD indices are assessed in the cath lab, namely on a branch-by-branch basis. However, in this study, CIPs were extracted over the entire angiographic view, reflecting the global contrast dynamics across all branches of the left coronary artery tree. Therefore, instead of calculating CMD indices for each individual branch, average values of IMR and CFR across four major branches within the LCA tree were obtained to represent overall microvascular health and its correlation with CIP:
\begin{equation}
\label{eqn:imr}
\text{IMR} = \frac{1}{N} \sum_{i=1}^{N}\text{IMR}_{i},
\end{equation}
\begin{equation}
\label{eqn:cfr}
\text{CFR} = \frac{1}{N} \sum_{i=1}^{N}\text{CFR}_{i},
\end{equation}

\subsection{Development of Data-driven Models}

\begin{figure}[h]
\centering
\includegraphics[width=0.8\textwidth]{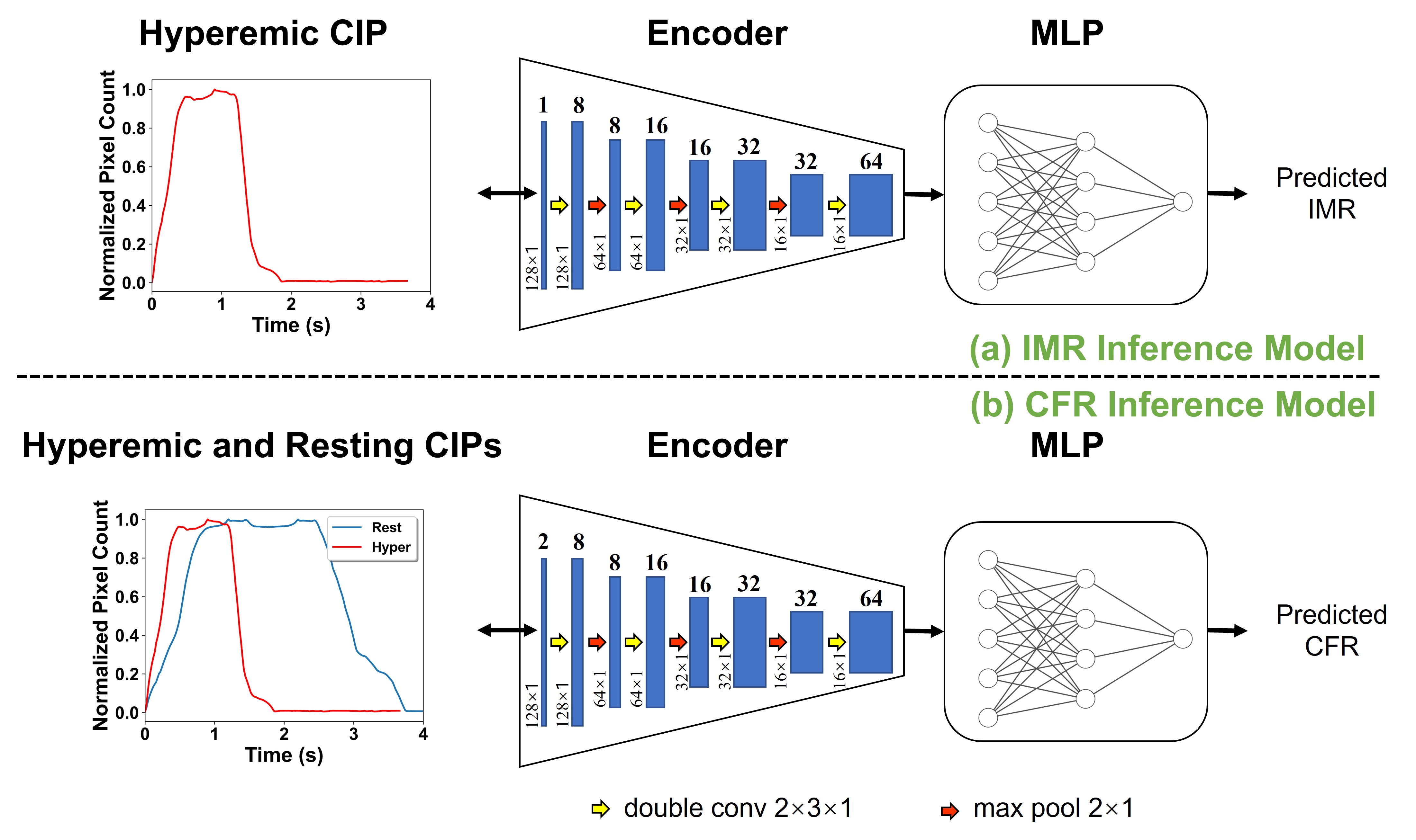}
\caption{\centering Architecture of data-driven models for (a) IMR and (b) CFR inference.}
\label{fig:ML_model}
\end{figure}

Data-driven models were developed to unveil the relationship between CIPs and CMD indices. To account for the distinct characteristics of different functional metrics of CMD, distinct model architectures were designed for each metric. For IMR prediction, a single-input-channel encoder–MLP architecture was employed, utilizing the CIP under hyperemic conditions as input, as illustrated in Fig. \ref{fig:ML_model} (a). In contrast, CFR prediction leveraged a dual-input-channel encoder–MLP architecture that processes both resting and hyperemic CIPs, as shown in Fig. \ref{fig:ML_model} (b). These metric-specific architectures are essential to accommodate the distinct physiological features of each CMD index.

The encoders in both architectures were designed to extract hierarchical features from the CIPs using a sequence of convolutional and max-pooling layers. Convolutional layers, with a kernel size of $2 \times 1$ and stride 1, were selected to detect fine-grained temporal patterns, while max-pooling layers with a $2 \times 1$ kernel and stride 2 were used to progressively reduce dimensionality. The number of channels, representing the number of kernels in each layer, is indicated above each block in Fig. \ref{fig:ML_model}, and the corresponding feature map dimensions are shown alongside the blocks. As the encoder advances through successive layers, the number of channels increases to enhance the capacity of the model for complex feature extraction, while the dimensions decrease to condense the temporal information into a compact representation that facilitates efficient downstream processing by the multi-layer perceptrons (MLP). 

Following the encoders, the extracted feature vectors are fed into MLPs to predict the CMD indices. For both IMR and CFR predictions, the MLP architecture consists of four fully connected layers with 1024, 1024, 128, and 1 neurons. As the number of neurons decreases through the MLP layers, the network transforms high-dimensional feature representations into increasingly abstract and compact forms, ultimately enabling precise regression of the target CMD indices.

\begin{figure}[H]
\centering
\includegraphics[width=0.7\textwidth]{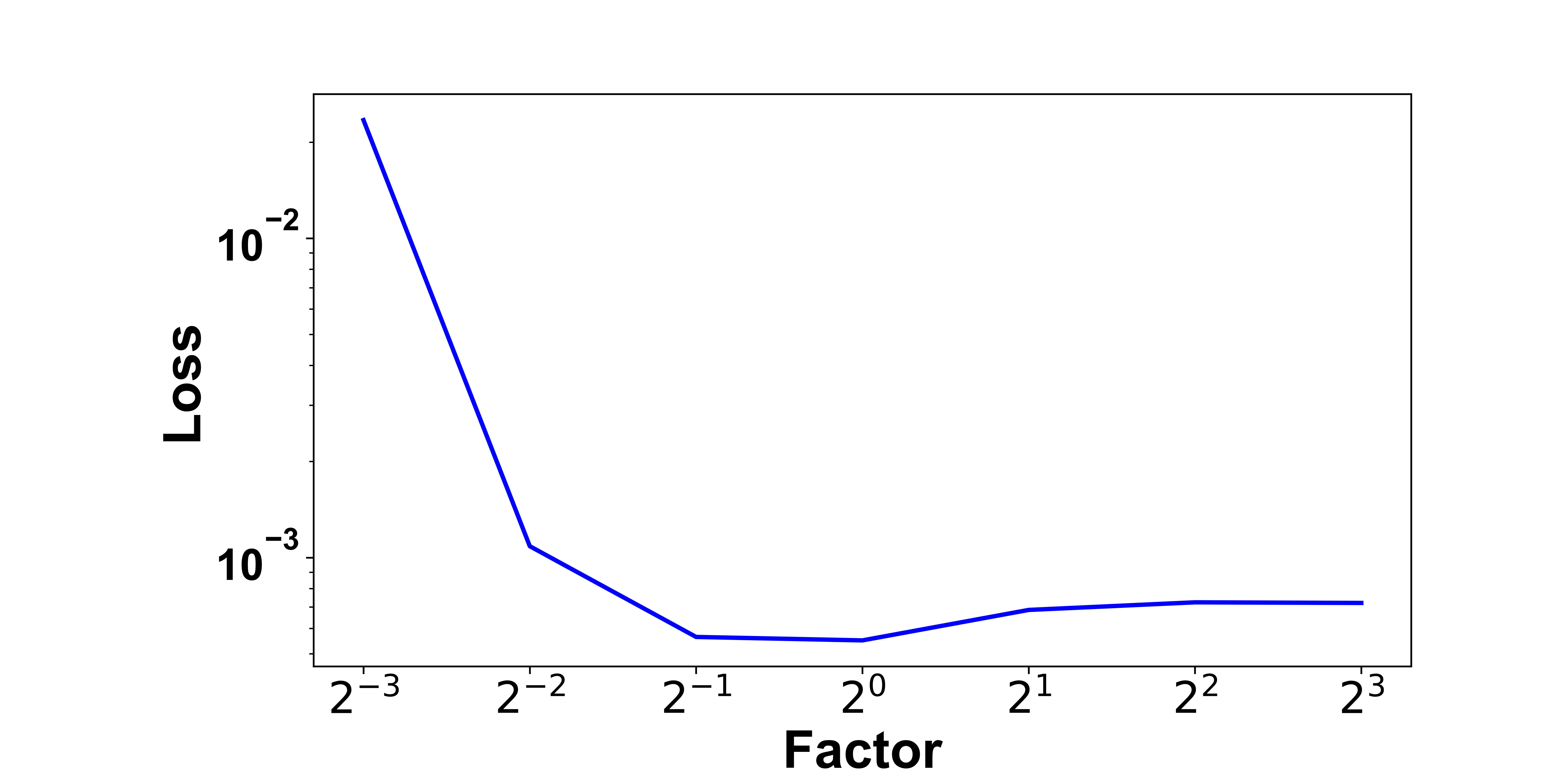}
\caption{\centering Validation losses for different architecture sizes.}
\label{fig:HyperOP}
\end{figure}

The size of the data-driven model architectures, including the number of encoder channels and MLP neurons, was selected through a comprehensive hyperparameter search to maximize model accuracy while avoiding overfitting. For IMR inference, this process involved evaluating a series of scaled versions of the design shown in Fig. \ref{fig:ML_model} (a), using scaling factors of [1/8, 1/4, 1/2, 1, 2, 4, 8]. Each configuration was evaluated using K-fold cross-validation, and the corresponding validation losses are presented in Fig. \ref{fig:HyperOP}. Among these configurations, the data-driven model with a scaling factor of 1 achieved the lowest cross-validation loss and was therefore selected as the final size of the architecture for IMR inference. The same architecture size was applied to CFR inference, given the similarity in input data between the two data-driven models. However, an independent hyperparameter search could be conducted for the CFR model to explore task-specific optimizations.

\subsection{Uncertainty Quantification}

Data-driven models serve as approximate surrogates for complex physical systems and are therefore subject to uncertainties in their predictions. These uncertainties can be broadly categorized into two types: \textit{aleatoric} and \textit{epistemic}~\cite{hullermeier2021aleatoric, kendall2017uncertainties, abdar2021review, yang2023neural, yang2024attention}. Aleatoric uncertainty originates from intrinsic randomness or noise present in the data. As it reflects variability inherent to the data generation process, this type of uncertainty is generally irreducible, regardless of the amount of available data. In this work, the data were generated from multi-physics simulations using stringent convergence criteria. As a result, the influence of numerical noise is minimal, and aleatoric uncertainty is considered negligible.

In contrast, \textit{epistemic uncertainty} arises from a lack of knowledge about the system, often due to model inadequacies, limited data availability, or inadequate representation of the input space. Unlike aleatoric uncertainty, epistemic uncertainty is theoretically reducible through additional data collection, refined model sizes, or better experimental designs. In the context of data-driven models, epistemic uncertainty is typically estimated by evaluating the variability across predictions. Several techniques have been developed for this purpose, including Bayesian neural networks, Monte Carlo dropout, and deep ensemble methods~\cite{zhang2021bayesian,zhang2023label,hullermeier2021aleatoric, kendall2017uncertainties, lakshminarayanan2017simple, gal2016dropout, ganaie2022ensemble}. In this study, a deep ensemble approach is adopted to quantify epistemic uncertainty. Specifically, an ensemble of $M = 5$ independently trained models is constructed, and the variability in their predictions is used to estimate uncertainty. The epistemic uncertainty, denoted by $\sigma^2_{\text{epistemic}}$, is calculated as:

\begin{equation}
\begin{aligned}
&   \left\{
\begin{aligned}
&\sigma^2_{\text{epistemic}} = \frac{1}{M} \sum_{j=1}^{M}(\hat{y}_{j}-\hat{\mu})^{2} \\
&\hat{\mu} = \frac{1}{M} \sum_{j=1}^{M}\hat{y}_{j}
\end{aligned}
\right. 
\end{aligned}
\end{equation}
where $\hat{\mu}$ represents the mean prediction across all ensemble members $\hat{y}_{j}$.

\section{Results}\label{Results}

This section first introduces two reference simulations under resting and hyperemic conditions to demonstrate the physiological relevance of the multi-physics model results used for machine learning training. Subsequently, inference results of IMR and CFR are presented to demonstrate the feasibility of the proposed framework.

\subsection{Multi-physics Simulation Results for Reference Cases}
\label{referencecases}

\begin{figure}[H]
\centering
\includegraphics[width=1\textwidth]{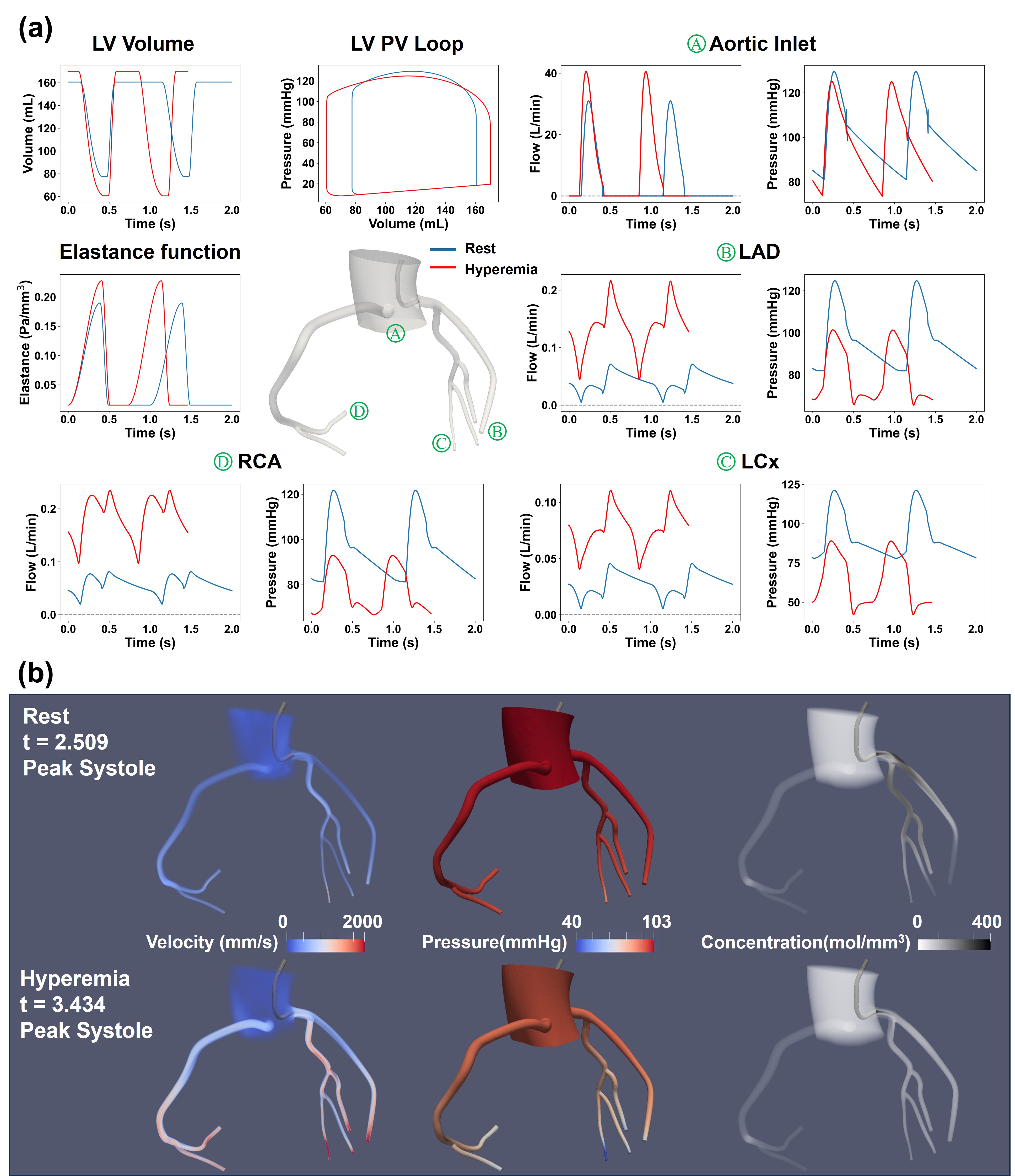}
\caption{\centering Simulation results of the reference hemodynamic states under rest and hyperemia: (a) cardiac function and coronary artery hemodynamics. (b) CFD contour plots of velocity, pressure, and contrast agent concentration at peak systole for rest (top row) and hyperemia (bottom row).}
\label{fig:Hemodynamics}
\end{figure}

Fig. \ref{fig:Hemodynamics} displays results of the reference simulations for both resting and hyperemic conditions, representing a healthy individual. Panel (a) shows cardiac function hemodynamics, including left ventricular (LV) volume, elastance function, and pressure-volume (PV) loop, along with flow and pressure waveforms at aortic inlet, LAD, LCx, and RCA. Panel (b) displays contour plots of velocity, pressure, and contrast agent concentration at peak systolic flow, at times $t_\text{rest} = 2.509$ s for rest and $t_\text{hyper} = 3.434$ s for hyperemia.

The coronary artery flow waveforms exhibit physiologically realistic patterns, characterized by a diastolic-dominant flow under both resting and hyperemic conditions. Key hemodynamic indices are summarized in Table \ref{tab:Hemo_numerical}. From resting to hyperemic states, there is a decrease in cardiac cycle length of 0.27 s, a 9.443 ml increase in end-diastolic volume ($\text{EDV}$), a 16.947 ml reduction in end-systolic volume ($\text{ESV}$), a 26.390 ml increase in stroke volume (SV), and a 12.6\% improvement in ejection fraction (EF). These changes resulted in an 80.456\% increase in cardiac output ($Q_{\text{mean}}$) and reductions of 4.483 mmHg in systolic aortic pressure ($P_{\text{sys}}$) and 7.413 mmHg in diastolic aortic pressure ($P_{\text{dia}}$). Furthermore, these results yield coronary flow reserve (CFR) of 3.097 and 3.161 in the left and right coronary trees (LCT and RCT), respectively, indicating a relatively healthy coronary function.

\begin{table}[H]
\centering
\footnotesize
\renewcommand{\arraystretch}{1.5} 
\setlength{\tabcolsep}{5pt}
\caption{\centering Hemodynamic metrics from calibrated simulations under resting and hyperemic conditions.}
\label{tab:Hemo_numerical}
\resizebox{\textwidth}{!}{  
\begin{tabular}{c c c c c c c c c c c}
\hline
  & $Q_{\text{mean}}$ & $Q_{\text{max}}$ & $P_{\text{sys}}$  & $P_{\text{dia}}$ & $\text{EDV}$& $\text{ESV}$& SV & EF& $Q_\text{LCT}$& $Q_\text{RCT}$\\ 
  & (L/min) & (L/min) & (mmHg) & (mmHg) & (ml) & (ml)  & (ml) &  &(L/min)& (L/min)\\
\hline
Rest & 4.982& 30.970& 129.527& 81.070& 160.608& 77.563& 83.045& 51.707\%& 0.145&  0.097
\\ 
 Hyperemia & 8.990& 40.535& 125.045& 73.657& 170.051& 60.616& 109.435& 64.354\%& 0.449& 0.307
\\ 
\hline
\end{tabular}
}
\end{table} 

\begin{figure}[h]
\centering
\includegraphics[width=0.5\textwidth]{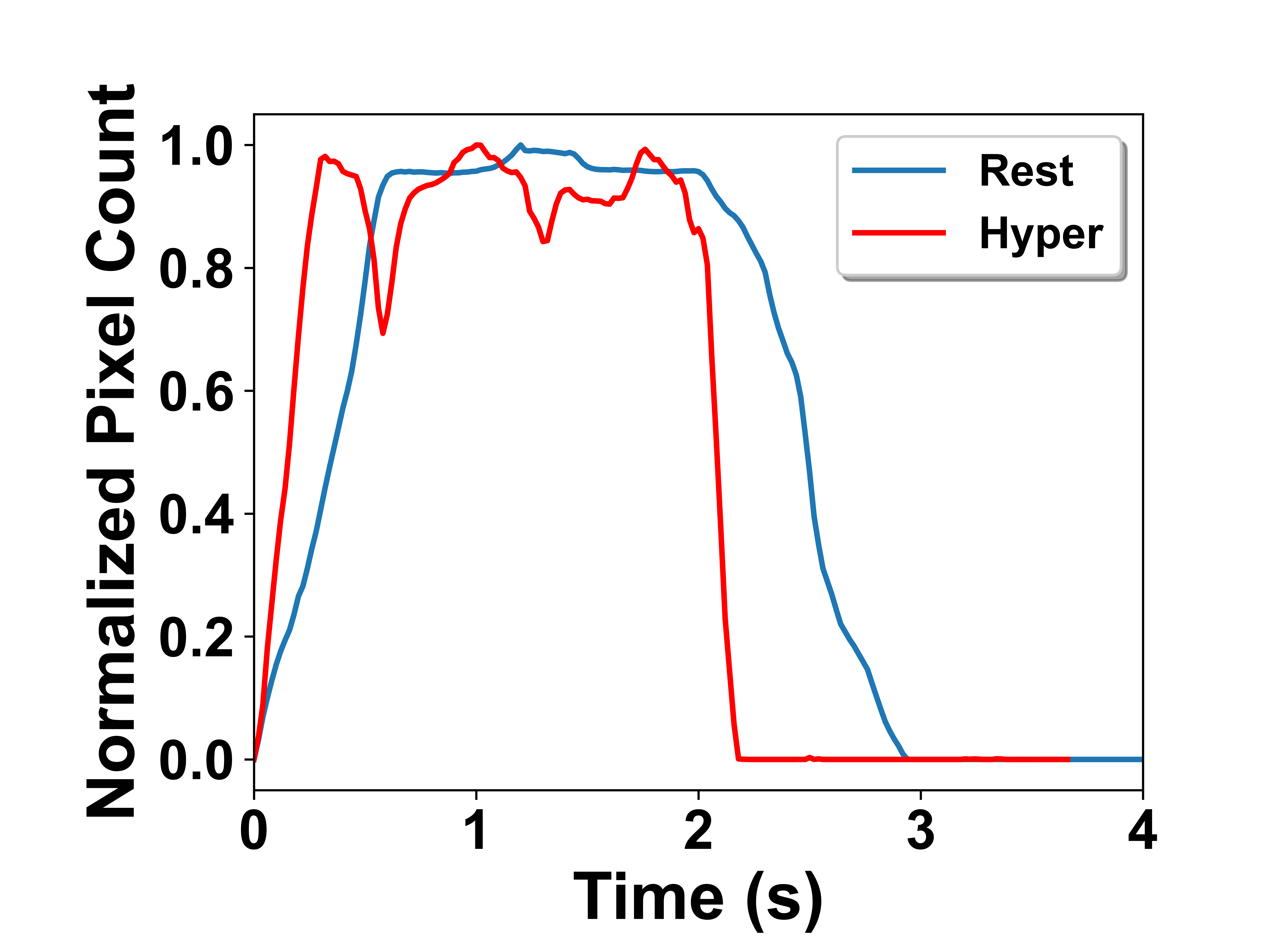}
\caption{\centering CIP for rerefence resting and hyperemic conditions.}
\label{fig:CIP_comparison}
\end{figure}

From these results for contrast agent dynamics, a process of projection, thresholding and pixel count is used to generate CIP, as sketched in Fig \ref{fig:Framework} and explained in detail elsewhere\cite{yang2024multi}. Fig. \ref{fig:CIP_comparison} shows the CIP for the reference resting and hyperemic conditions. These were obtained assuming RAO = $21.9^\circ$ and CAU = $18.3^\circ$ projection angles. Under hyperemic conditions, the CIP exhibits a noticeably quicker contrast injection and washout profile, characterized by steeper rising and falling slopes.

\subsection{IMR Inference} \label{IMR_Inference}

\begin{figure}[h]
\centering
\includegraphics[width=0.6\textwidth]{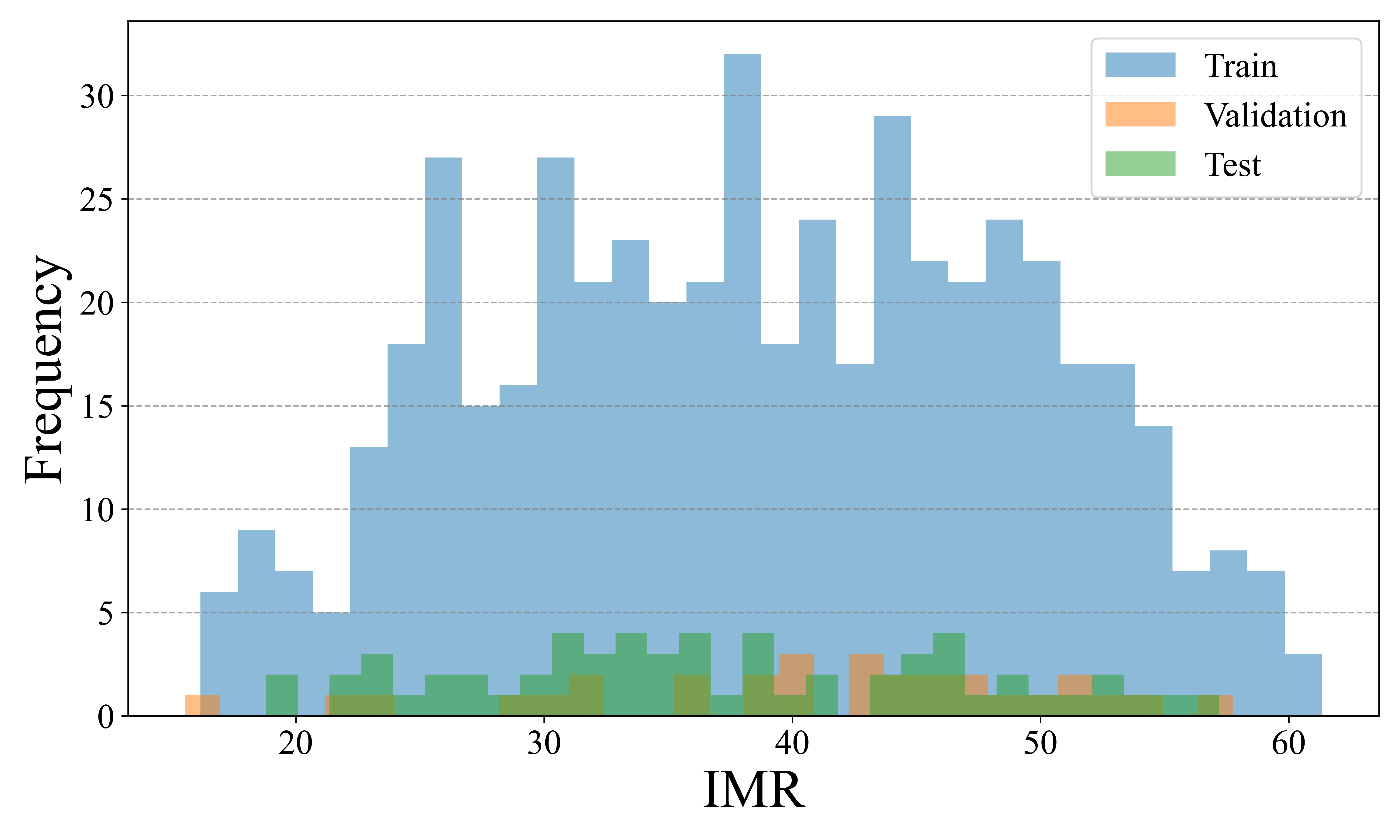}
\caption{\centering Distribution of IMR values.}
\label{fig:IMR_distribution}
\end{figure}

Based on 600 multi-physics simulations under hyperemia generated as described in \ref{DataGeneration}, 600 data pairs were defined, each consisting of a CIP and the corresponding IMR. These data pairs were used to construct the data-driven model for IMR inference.  This dataset was split into three subsets: 85\% of the data was used for training, 5\% for validation during the training process to prevent overfitting and guide hyperparameter tuning, and the remaining 10\% was reserved for testing the data-driven model performance on unseen data. The distribution of IMR values across the dataset is illustrated in Fig. \ref{fig:IMR_distribution}.

\begin{figure}[H]
\centering
\includegraphics[width=0.5\textwidth]{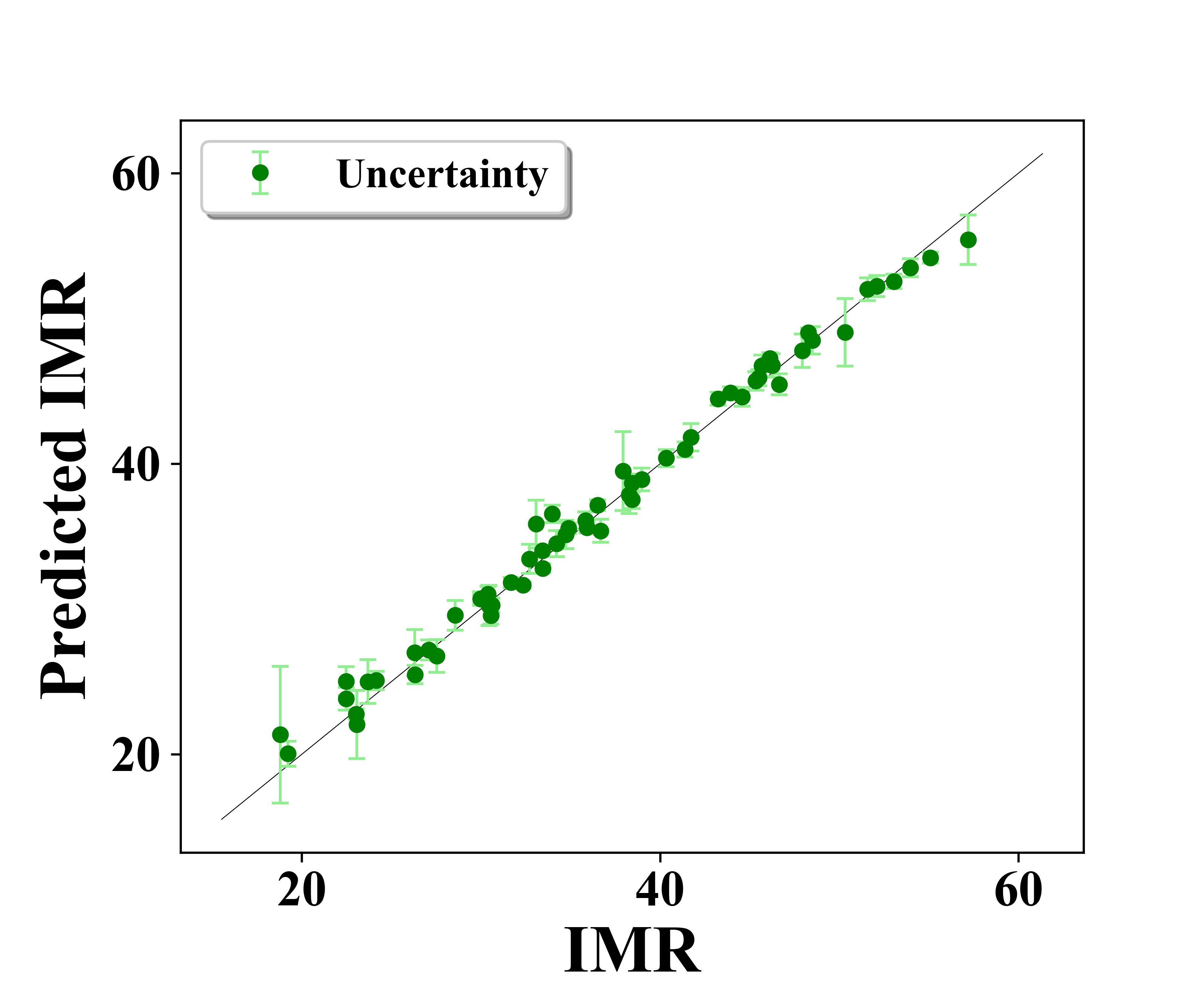}
\caption{\centering IMR inference for testing dataset.}
\label{fig:IMR_Inference}
\end{figure}

Fig.~\ref {fig:IMR_Inference} presents a comparison between the IMR values predicted by the data-driven model and ground truth values obtained from multi-physics simulations on the testing dataset. Additionally, the figure displays the 95\% confidence intervals (defined as $\pm 1.96\sigma_\text{epistemic}$) representing the epistemic uncertainty associated with the model predictions. Table~\ref{tab:IMR_MSE} summarizes the quantitative performance of the model in terms of mean squared error (MSE), coefficient of determination ($R^2$), and estimated epistemic uncertainty ($\sigma^2_{\text{epistemic}}$). The results indicate strong agreement between the model predictions and simulation-derived IMR values across a broad range of IMR values (20–60), with an MSE of 1.015 and an $R^2$ of 0.989, demonstrating both accuracy and robustness of the data-driven inference approach. 

\begin{table}[H]
\centering
\small
\caption{\centering Performance metrics of the data-driven model on the testing dataset.}
\label{tab:IMR_MSE}
\begin{tabular}{c c c c}
\hline
Metric &   MSE&   $R^2$&   $\sigma^2_{\text{epistemic}}$\\ \hline
Value & 1.015   & 0.989  &  0.446 \\
\hline
\end{tabular}
\end{table}

\begin{figure}[H]
\centering
\includegraphics[width=0.5\textwidth]{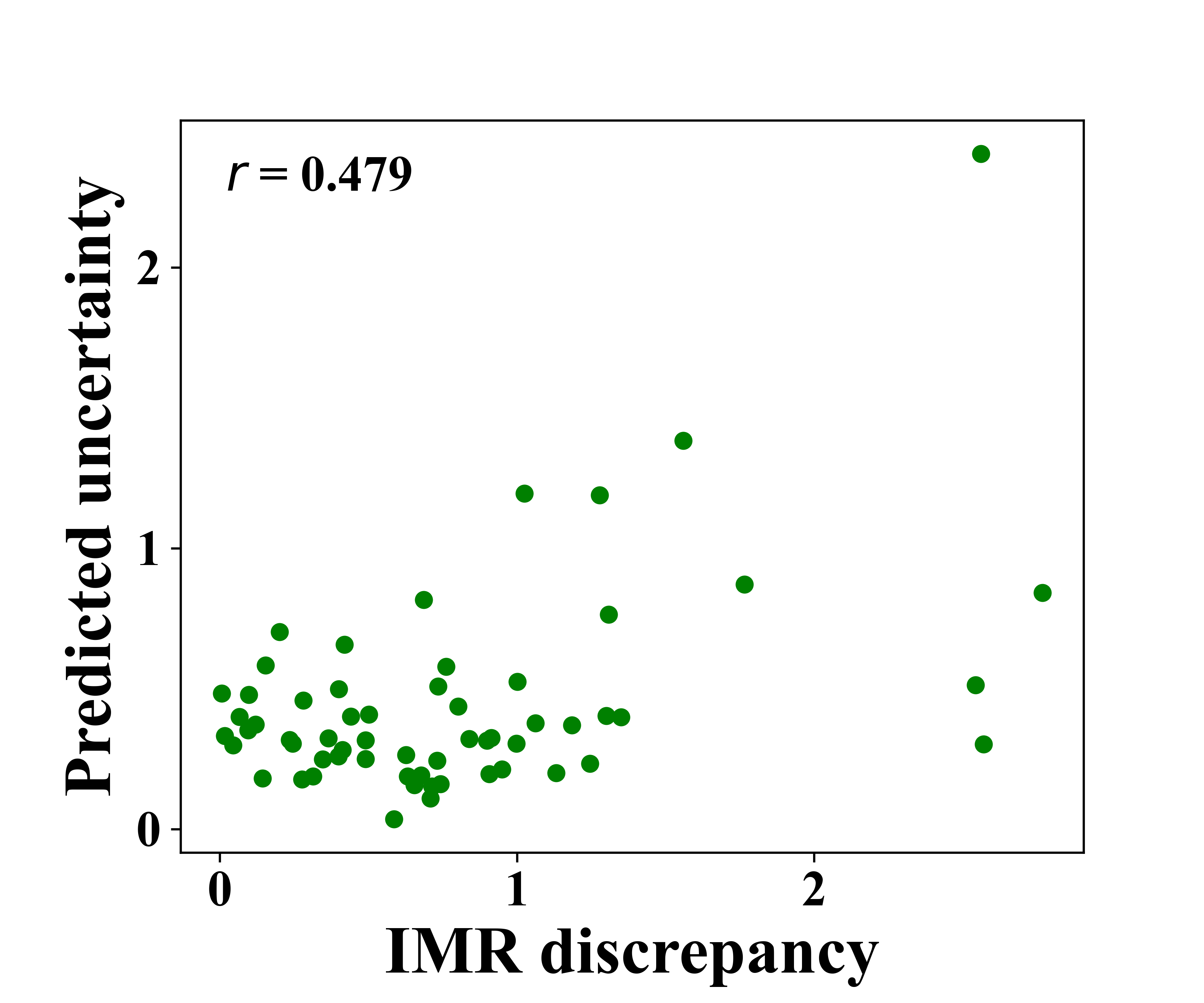}
\caption{\centering Scatter plot of model-predicted uncertainty vs IMR discrepancy on the testing dataset.}
\label{fig:IMR_discrepancy_uncertainty}
\end{figure}

Fig.~\ref{fig:IMR_discrepancy_uncertainty} displays a scatter plot illustrating the relationship between the IMR predicted discrepancy (i.e., the absolute value of the difference between model-predicted IMR and ground truth IMR) and the corresponding epistemic uncertainty $\sigma_{\text{epistemic}}$. A positive correlation is observed, with a Pearson correlation coefficient of $r = 0.479$, indicating that higher prediction discrepancies tend to be associated with greater uncertainty estimates. This highlights the capability of the proposed uncertainty quantification approach to reliably reflect the confidence of the model in its predictions.

\subsection{CFR Inference} 

\begin{figure}[h]
\centering
\includegraphics[width=0.6\textwidth]{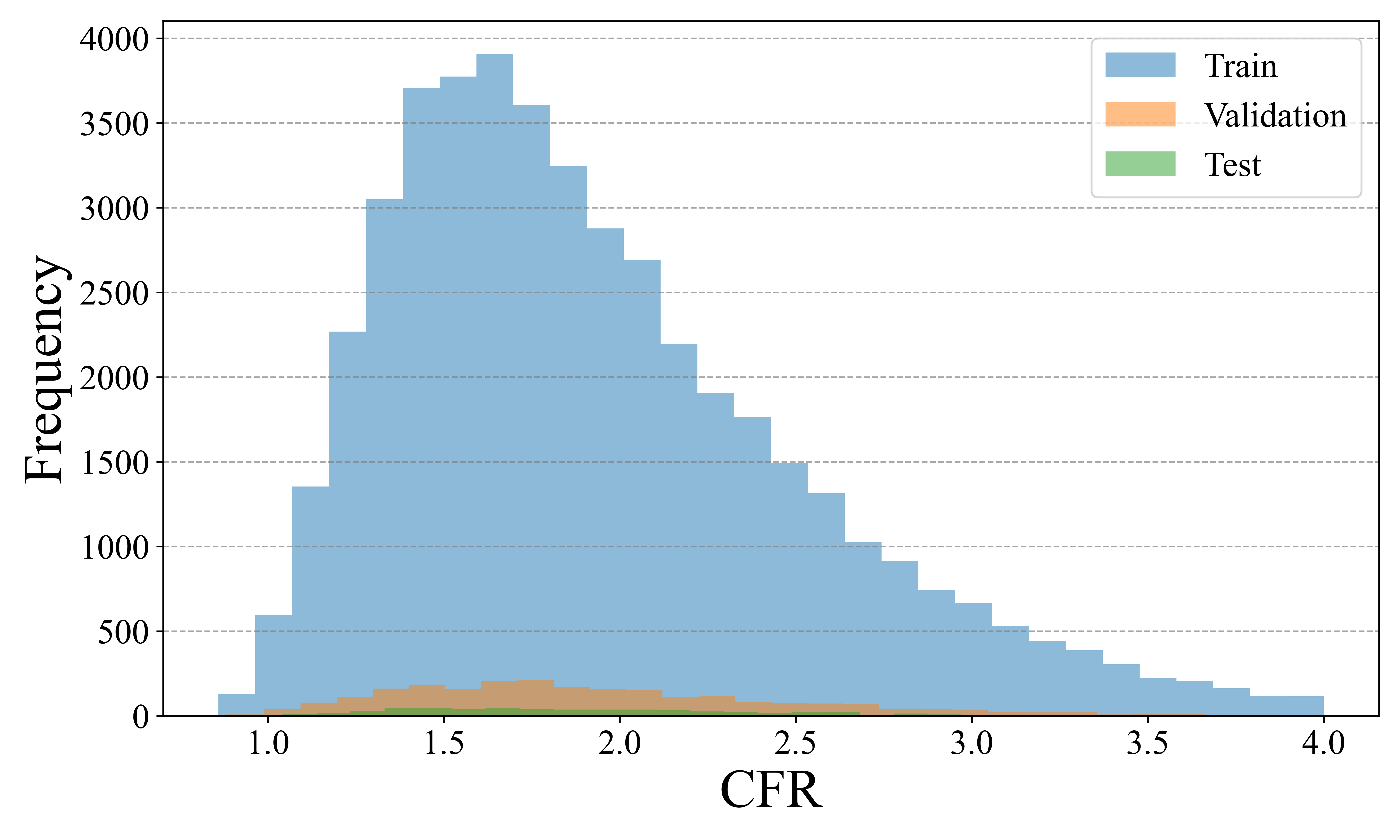}
\caption{\centering Distribution of CFR values.}
\label{fig:CFR_distribution}
\end{figure}

For the CFR inference task, each data point was constructed from a pair of simulations conducted under hyperemic and resting conditions, producing two CIPs and a corresponding computed CFR value. Out of 600 hyperemic and 100 resting multi-physics simulations, 540 hyperemic and 90 resting cases were used for model training and validation. These were then combined to generate 48{,}600 data pairs. Out of the remaining 60 hyperemic and 10 resting simulations, 600 data pairs were formed for model testing. To ensure physiological relevance, data pairs yielding CFR values larger than 4, which correspond to extremely healthy individuals without heart conditions \cite{bourdarias1995coronary}, were excluded. After removing cases with CFR larger than 4, 48{,}138 training/validation pairs and 598 testing pairs were obtained. Training and validation sets were split with a 9:1 ratio. Fig. \ref{fig:CFR_distribution} shows the distribution of CFR values, which are intentionally concentrated around 2 to improve the predictive performance near the diagnostic threshold. A CFR value around 2 has been used as a reference threshold for CMD diagnosis \cite{johnson2012discordance,kim2022differential,joh2022prognostic,murai2020coronary}, although the exact cutoff may vary depending on the measurement technique and clinical context \cite{meuwissen2009role,ang2022phenotype}.

\begin{figure}[h]
\centering
\includegraphics[width=0.5\textwidth]{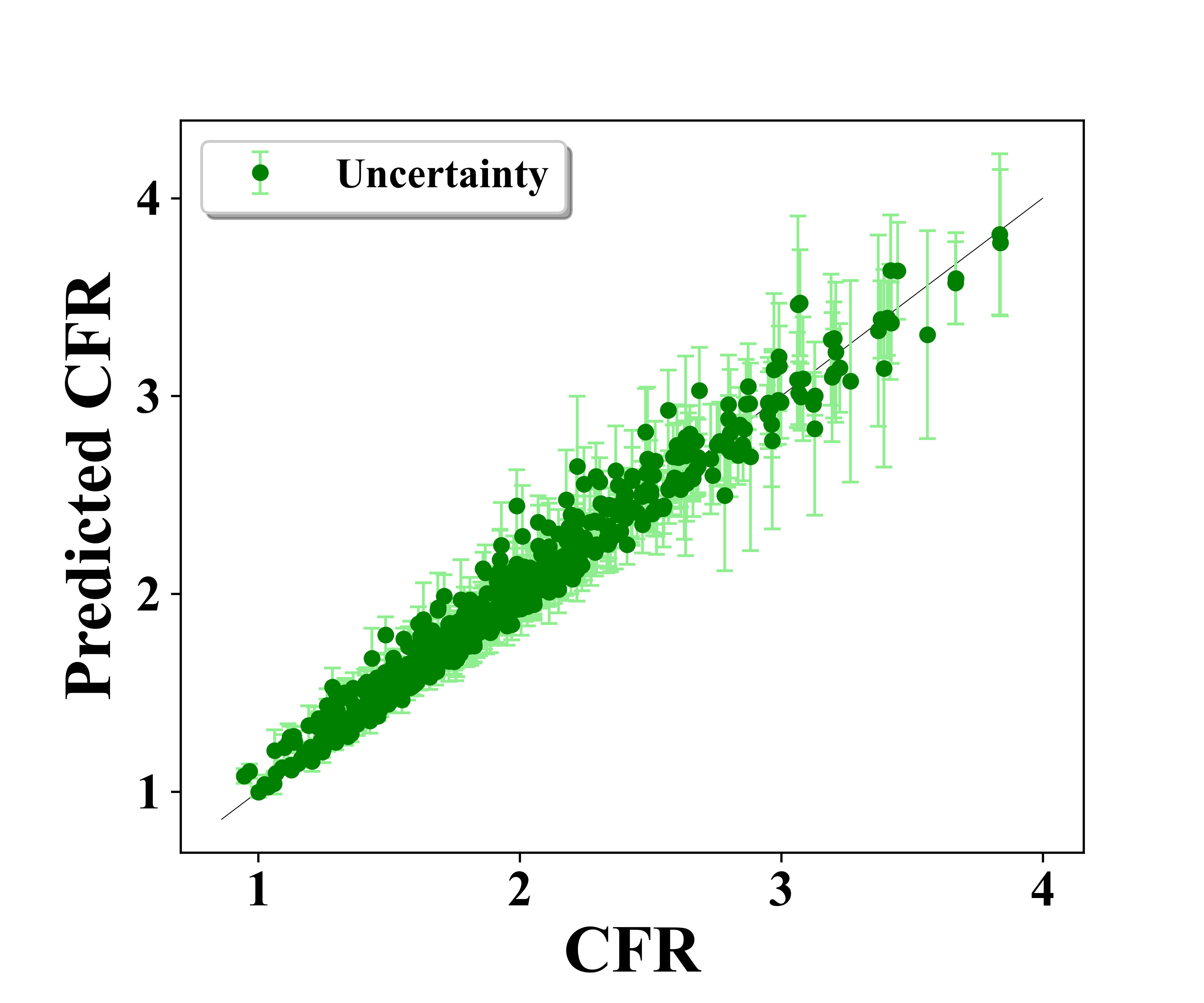}
\caption{\centering CFR inference for testing dataset.}
\label{fig:CFR_Inference}
\end{figure}

Fig.~\ref{fig:CFR_Inference} illustrates a comparison between CFR values predicted by the data-driven model and those obtained from the multi-physics simulations on the testing dataset. The 95\% confidence intervals, calculated as $\pm 1.96\sigma_\text{epistemic}$, are also shown to reflect the epistemic uncertainty associated with the predictions. Table~\ref{tab:CFR_MSE} reports the quantitative performance of the model on the testing dataset, including the mean squared error (MSE), coefficient of determination ($R^2$), and estimated epistemic uncertainty ($\sigma^2_{\text{epistemic}}$). The results demonstrate a strong correlation between predicted and simulated CFR values over the (1–4) range, with an MSE of 0.010 and an $R^2$ of 0.967, highlighting the accuracy of the proposed data-driven model. 

\begin{table}[H]
\centering
\small
\caption{\centering Performance metrics of the data-driven model on the testing dataset.}
\label{tab:CFR_MSE}
\begin{tabular}{c c c c}
\hline
Metric &   MSE&   $R^2$&   $\sigma^2_{\text{epistemic}}$\\ \hline
Value & 0.010   & 0.967  &  0.052 \\
\hline
\end{tabular}
\end{table}

\begin{figure}[H]
\centering
\includegraphics[width=0.5\textwidth]{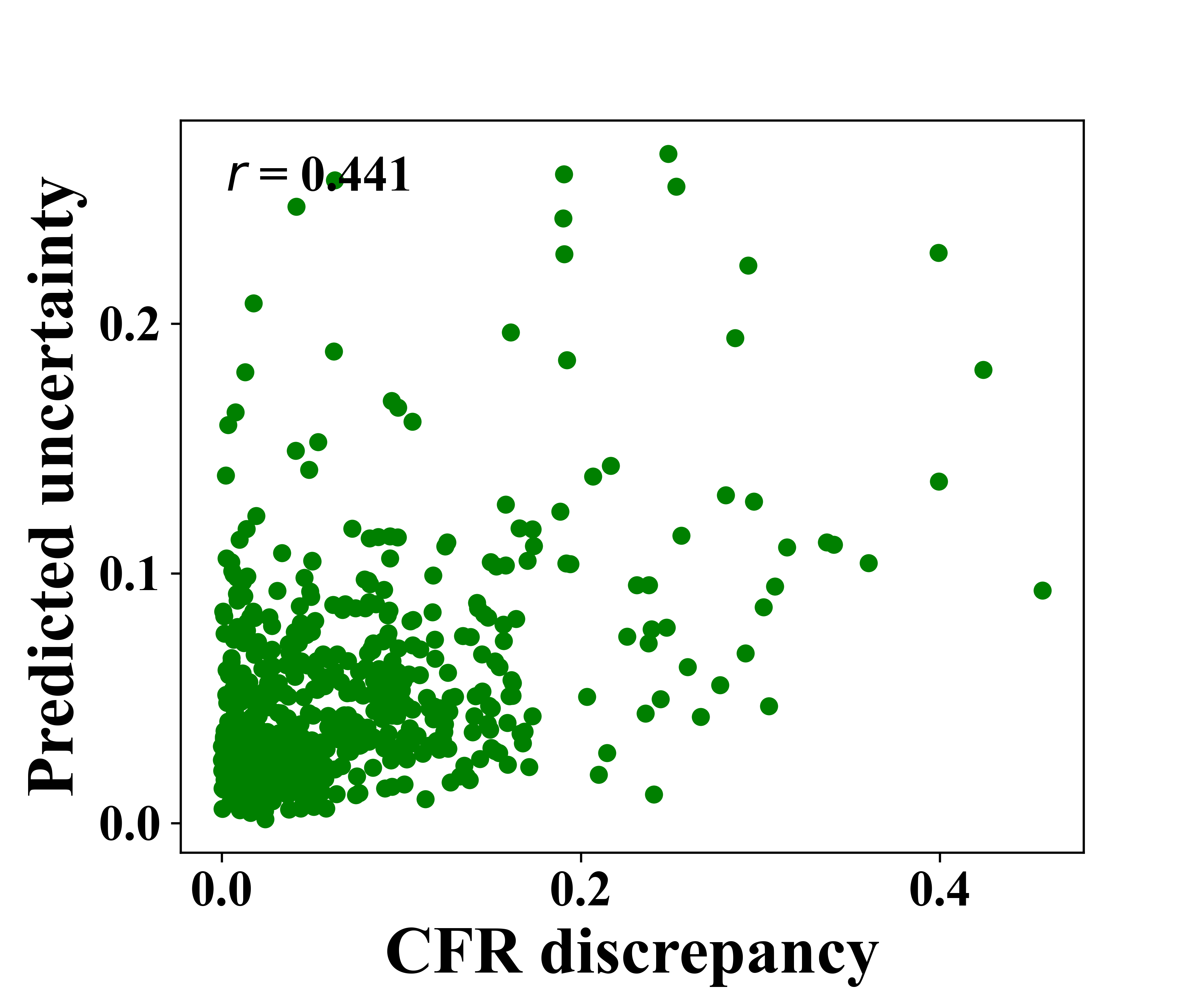}
\caption{\centering Scatter plot of model-predicted uncertainty versus CFR discrepancy on the testing dataset.}
\label{fig:CFR_discrepancy_uncertainty}
\end{figure}

Fig.~\ref{fig:CFR_discrepancy_uncertainty} displays a scatter plot illustrating the relationship between CFR discrepancy and the corresponding predicted uncertainty $\sigma_{\text{epistemic}}$. A positive Pearson correlation coefficient of $r = 0.441$ is obtained. Consistent with observations for IMR inference, predictions with greater discrepancy relative to ground truth tend to exhibit higher uncertainty, further validating the capability of the proposed method to capture the confidence of the model and inform trustworthiness in its predictions.

\section{Discussion}\label{Discussion}

The results presented here demonstrate the effectiveness and physiological relevance of the proposed data-driven framework for inferring key CMD indices from computational angiography. This section discusses the performance of the framework, and outlines limitations. Furthermore, we also discuss expansions of the current framework to cases with simultaneous epicardial and microvascular disease.

\subsection{Framework Performance}
The reference multi-physics simulations under resting and hyperemic conditions presented in Section \ref{referencecases} exhibit physiologically relevant behavior, with ventricular and coronary flow patterns aligning well with characteristics of a healthy subject. Notably, the observed diastolic-dominant flow in the coronary arteries and the computed CFR values (3.097 in LCT and 3.161 in RCT) fall within expected physiological ranges. A central component of the proposed framework is the use of the CIP as a signal for describing coronary hemodynamics. Easily extractable from angiographic image sequences, CIPs reflect the spatiotemporal evolution of contrast agent through the coronary vasculature and inherently encode information on flow dynamics, vessel resistance, and microcirculatory function. Moreover, since CIP generation requires only conventional contrast injection and X-ray acquisition, this signal can be obtained within existing clinical workflows. Results show that CIPs under hyperemic conditions exhibit steeper upslope and downslope patterns compared to rest, consistent with faster contrast transit and washout. The physiologically meaningful flow and pressure waveforms, and the characteristics of the synthetic CIPs indicate that the multi-physics model of contrast injection is a good foundation for generating synthetic data to support the construction of data-driven models for CMD indices.

An encoder plus MLP architecture for IMR and CFR inference was considered. A hyperparameter search was performed to identify the optimal size of the architecture (see Fig. \ref{fig:HyperOP}). Results suggest that, as long as the model size and architecture are within a reasonable range, the overall performance is not highly sensitive to model size. The data-driven models trained for IMR and CFR inference achieved high predictive accuracy when evaluated against physics-based synthetic datasets (see Tables \ref{tab:IMR_MSE} and \ref{tab:CFR_MSE}). Additionally, epistemic uncertainty estimation enables a robust assessment of model confidence. For both IMR and CFR predictions, a positive correlation was observed between prediction errors and estimated uncertainty, reinforcing the interpretability and reliability of the proposed framework, an essential aspect for clinical applications where understanding the trustworthiness of model outputs is critical. In some cases, the uncertainty intervals overlap with the diagnostic threshold. These intervals, shown in Fig.~\ref{fig:CFR_Inference}, are not intended as binary decision criteria but rather as confidence ranges that quantify prediction reliability. Such uncertainty-aware outputs enhance interpretability and provide clinicians with an understanding of how confident the model is in each prediction, especially in borderline scenarios.

Additionally, different CFR thresholds are used in clinical practice. For instance, several studies have adopted CFR = 2 as the diagnostic threshold \cite{johnson2012discordance,kim2022differential,joh2022prognostic,murai2020coronary}, while CFR = 2.5 is also widely employed for CMD diagnosis \cite{rahman2020physiological,toya2021risk,writing20212021}. Nonetheless, we submit that there is no universally fixed threshold for CMD diagnosis—the critical value should be interpreted within a range (2-2.5), as also indicated by the Cleveland Clinic (\url{https://my.clevelandclinic.org/health/diagnostics/24028-coronary-flow-reserve}). If CFR = 2.5 had been adopted as the diagnostic threshold, a correspondingly adjusted synthetic data distribution (Fig.~\ref{fig:CFR_distribution}) would have been generated.

\subsection{Limitations}
Several limitations should be acknowledged to contextualize the findings and guide future work. First, the synthetic data utilized here was based on a single patient-specific coronary geometry. Despite the physiologically relevant results of the multi-physics model, this single geometric model choice limits the generalizability to broader patient populations with anatomical variability. 

Second, the data-driven models were trained and validated exclusively on synthetic datasets generated through the multi-physics model. Although the multi-physics model was calibrated to match available physiologic data, it lacks certain aspects of the complexity, variability, and noise commonly encountered in clinical angiographic data. In particular, the presented pipeline for generating computational angiograms does not incorporate dynamic factors present in real-world imaging, such as table motion, respiratory and cardiac motions, vessel compliance, and operator-dependent variations in acquisition protocols. 

Third, the size of our design space is limited, both with regard to the parametrization of the microvascular disease as well as the contrast injection protocol. All coronary branches were assumed to exhibit similar levels of microvascular dysfunction, with LPM scaled uniformly across branches. In fact, microvascular disease severity is often heterogeneous. Similarly, contrast injection protocols were kept constant for all simulations, with fixed injection duration and flow rate (see Fig. \ref{fig:Catheter_flow}). However, injection characteristics vary substantially and can influence the temporal and spatial distribution of contrast, thereby affecting the observed CIP dynamics. While expanding the size of the design space will improve the generalizability of the data-driven model, the number of multi-physics simulations required to resolve the design space will accordingly be much larger. 

Fourth, while IMR and CFR indices are acquired for individual vessels in clinical practice, in this study, we estimated average indices for all vessels of the tree. Therefore, we are providing a characterization of global microvascular health, rather than individual branches, potentially leading to misleading interpretations when heterogeneity is present across branches. However, the approach presented here could be easily adapted to single-vessel analysis. 

Fifth, a single fixed angiographic projection angle was used for all simulations. Although this simplification reduces computational complexity, it neglects the known sensitivity of CIP features to view angle. Relying on a single projection may introduce interpretive bias and compromise the robustness of the performance of the framework under varied clinical imaging conditions. We could describe the contrast washout as a 3D process, circumventing the need for projection of the simulation results onto 2D planes.
Given these limitations, we have not included clinical angiograms for model validation. This will be the subject of subsequent studies.

Lastly, from a clinical perspective, both CFR and IMR indices have been shown to display limited reproducibility and operator dependence when using techniques such as Doppler or bolus thermodilution \cite{everaars2018doppler,gallinoro2023reproducibility}. The quality of these indices could therefore affect the development of machine learning models. However, continuous thermodilution has recently demonstrated improved stability and reproducibility for CMD assessment \cite{candreva2021basics,gallinoro2023reproducibility}.

\subsection{Simultaneous Consideration of Epicardial and Microvascular Disease}

A significant assumption of this work lies in the lack of epicardial disease. In reality, many diagnostically relevant cases involve a combination of epicardial and microvascular disease, and CFR is influenced by both. Therefore, the machine learning models presented here must ultimately be aware of potential epicardial disease. Furthermore, the multi-physics model must also be able to discern between epicardial and CMD contributions to CFR.

\begin{figure}[h]
\centering
\includegraphics[width=1\textwidth]{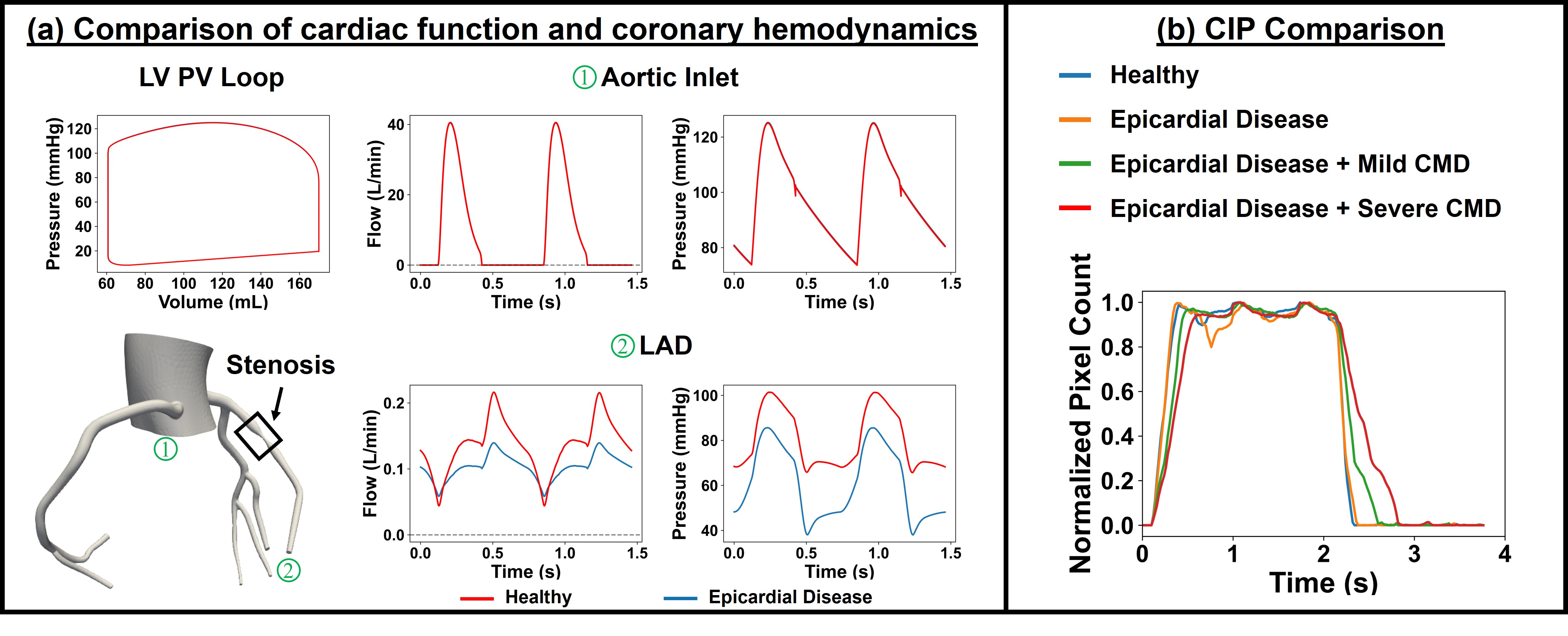}
\caption{\centering (a) Representative waveforms for the healthy subject and the patient with epicardial disease; (b) CIP comparison among the healthy subject, the epicardial disease–only case, and the epicardial disease cases with mild and severe CMD.}
\label{fig:Epicardial_Comparison}
\end{figure}

To partially address this limitation and to shed light on the suitability of the multi-physics model to generate meaningful synthetic data, we performed additional simulations incorporating epicardial disease. Specifically, we introduced a 90\% area reduction stenosis in the mid-LAD, which yielded an FFR = 8,973/12,491 = 0.72 and has no microvascular disease (e.g., the only difference with the healthy case is the stenosis; the coronary microvascular parameters remain unchanged). Based on this geometry, two additional simulations were conducted to represent epicardial disease combined with mild CMD (100\% increase in microvascular resistance) and severe CMD (200\% increase in microvascular resistance), respectively. Fig.~\ref{fig:Epicardial_Comparison} (a) illustrates representative waveforms for the healthy subject and the patient with epicardial disease, and Fig.~\ref{fig:Epicardial_Comparison} (b) shows the CIP comparison among the healthy subject, the epicardial disease–only case, and the epicardial disease cases with mild and severe CMD.

The comparison of the profiles reveals that the decay slopes of the CIP curves for the healthy and epicardial-only cases are similar. In contrast, the inclusion of CMD results in noticeably slower decay slopes. This trend is consistent with our findings of patients without epicardial disease and confirms that the CIP washout slope remains primarily dictated by microvascular resistance, even in the presence of epicardial stenosis. These results support two key findings: (1) our assumption of excluding cases with epicardial disease in the current study is reasonable, since moderate stenosis has minimal influence on overall contrast washout dynamics; and (2) the multi-physics model is capable of generating physiologically realistic angiography simulations in the presence of epicardial disease, demonstrating that the proposed framework can be extended to cases with combined epicardial and microvascular dysfunction.

\section{Conclusion and Future Work}\label{Conclusion}

This study presents a novel data-driven framework for inference of coronary microvascular dysfunction (CMD) indices, namely index of microcirculatory resistance (IMR) and coronary flow reserve (CFR), from computational angiography data. The framework consists of two stages: training and testing. During the training stage, synthetic datasets consisting of contrast intensity profiles (CIPs) and corresponding CMD indices were generated under resting and hyperemic conditions using a physiologically validated multi-physics model. These datasets were used to train two data-driven models: a single-input-channel encoder–MLP architecture for IMR inference using hyperemic CIPs, and a dual-input-channel encoder–MLP model for CFR inference based on paired resting and hyperemic CIPs. In the testing stage, these models predict CMD indices from CIPs with associated uncertainty estimates. 

Our results demonstrate that (1) the multi-physics model produces physiologically realistic hemodynamics and serves as a robust foundation for generating training data. (2)  The CIPs effectively encode important features of coronary flow dynamics relevant to CMD. (3) The trained data-driven models achieve strong predictive performance when evaluated against physics-based synthetic datasets and demonstrate reliable uncertainty quantification. (4) Positive correlations between prediction errors and epistemic uncertainty further support the prediction confidence of the data-driven models. Overall, the proposed framework enables accurate, efficient, and interpretable estimation of CMD indices from CIPs, supporting non-invasive, real-time coronary functional assessment.

The key innovations of this work are: (1) it is the first to develop data-driven methods for CMD assessment using dynamic contrast information from coronary angiography, a domain where prior research has primarily focused on diagnosing epicardial disease (e.g., FFR estimation), with little to no work addressing CMD inference; and (2) it incorporates uncertainty quantification, enabling confidence intervals for predicted indices to enhance interpretability and clinical trust.

Future work will focus on improving both the physiological fidelity and clinical applicability of the proposed framework through several key developments. First, we will expand the anatomical and physiological diversity represented in the synthetic dataset. In this study, population-average values for ventricular function and coronary flow were assumed for a representative healthy individual, and then a wide range of CMD conditions was generated synthetically. In future work, we will expand the synthetic dataset to cover a broader spectrum of disease categories, including varying degrees of epicardial stenosis, CMD, and their combined presentations. This will be achieved by incorporating literature-derived distributions of systolic and diastolic ventricular function (from Doppler echocardiography) and coronary flow (from invasive measurements such as continuous thermodilution or cardiac PET). At the same time, incorporating multiple patient-specific geometries will further increase anatomical diversity. Together, these developments will substantially increase the size and diversity of the training dataset, and thus enhance the representativeness and generalizability of the machine-learning model across clinically relevant disease phenotypes. Second, integrating real-world imaging artifacts, such as respiratory and cardiac motion, table displacement, and vessel compliance, into the simulation pipeline will help bridge the fidelity gap between synthetic and clinical angiograms. Third, future work will consider vessel-specific microvascular conditions by independently varying lumped parameter model (LPM) parameters across different coronary branches, thereby capturing the heterogeneity commonly observed in clinical populations. Similarly, a broader range of contrast injection protocols will be explored by varying injection duration, rate, and catheter positioning to simulate real-world variability and improve the robustness of the framework. Fourth, incorporating single-vessel CIP extraction and analysis will enable localized assessments of microvascular health, which is more aligned with clinical diagnostic practice. Lastly, extending the framework to accommodate multiple angiographic view angles will account for projection-dependent variations in CIP features and enhance the performance of the framework under diverse clinical acquisition protocols. As a critical next step, the incorporation of clinical angiograms and corresponding invasive IMR/CFR measurements will be pursued to fine-tune and validate the data-driven models under clinical conditions.

\section*{Declarations}



\textbf{Code availability}:

The code for the multi-physics simulation is available at www.crimson.software.

Source code can be accessed via https://github.com/carthurs/CRIMSONFlowsolver (CRIMSON Flowsolver) under the GPL v3.0 license, and https://github.com/carthurs/CRIMSONGUI (CRIMSON GUI).

The underlying code for data-driven model construction is not publicly available but may be made available to qualified researchers on reasonable request from the corresponding author.

\bibliography{sample}
\end{document}